\documentclass[9pt,twocolumn,letter]{article}

\usepackage[left=0.5in,right=0.5in,top=0.5in,bottom=1in]{geometry}  
\usepackage{graphics}
\usepackage{graphicx}
\usepackage[title]{appendix}

\usepackage{multibib}
\newcites{supp}{References}

\usepackage{mathtools,stmaryrd,soul}
\usepackage{amsmath,amssymb}
\usepackage{authblk}
\usepackage{xcolor}

\usepackage[utf8]{inputenc}

\usepackage{caption}

\newcommand{\beginsupplement}{%
        \setcounter{table}{0}
        \renewcommand{\thetable}{S\arabic{table}}%
        \setcounter{figure}{0}
        \renewcommand{\thefigure}{S\arabic{figure}}%
        \setcounter{section}{0}
        \renewcommand{\thesection}{S\arabic{section}}
     }

\date{}

\begin{document}

\title{Single-shot Digital Optical Fluorescence Phase Conjugation Through Forward Multiply Scattering Samples}

\author[1]{Tengfei Wu}
\author[1]{Yixuan Zhang}
\author[1]{Baptiste Blochet}
\author[1]{Payvand Arjmand}
\author[1,2,3]{Pascal Berto}
\author[1,3*]{Marc Guillon}

\affil[1]{Saints-Pères Paris Institute for the Neurosciences, CNRS UMR 8003, Université de Paris, 45 rue des Saints-Pères, Paris 75006, France}
\affil[2]{Sorbonne Université, CNRS, INSERM, Institut de la Vision, 17 Rue Moreau, 75012 Paris, France}
\affil[3]{Institut Universitaire de France, Paris, France}

\affil[*]{ marc.guillon@u-paris.fr}


\maketitle


\textbf{
Aberrations and multiple scattering in biological tissues critically distort light beams into highly complex speckle patterns. In this regard, digital optical phase conjugation (DOPC) is a promising technique enabling in-depth focusing. However, DOPC becomes challenging when using fluorescent guide-stars for four main reasons: The low photon budget available, the large spectral bandwidth of the fluorescent signal, the Stokes shift between the emission and the excitation wavelength, and the absence of reference beam preventing holographic measurement.
Here, we demonstrate the possibility to focus a laser beam through multiple-scattering samples by measuring speckle fields in a single acquisition step with a reference-free, high-resolution wavefront sensor. By taking advantage of the large spectral bandwidth of forward multiply scattering samples, Digital Fluorescence Phase Conjugation (DFPC) is achieved to focus a laser beam at the excitation wavelength while measuring the broadband speckle field arising from a micron-sized fluorescent bead. 
}

\section*{Introduction}

\paragraph{In-depth laser focusing.}

Laser light focusing inside or through aberrant and scattering sample is a major challenge which is of interest for in-depth imaging~\cite{Ntziachristos_NM_10}, photo-stimulation~\cite{Li_CB_23} or phototherapy~\cite{Halas_ACR_08}.
At depth larger than the scattering mean free path $\ell_s$ (typically $\simeq 100~{\rm \mu m}$ in brain tissues~\cite{jacques2013optical}) light is scrambled into a complex random-like speckle pattern. Imaging ability is thus critically altered by reduced signal to noise ratios, even under non-linear photo-excitation~\cite{theer2006fundamental}.

\paragraph{Adaptive optics and wavefront shaping.}

In-depth light control has then been addressed both by adaptive optics~\cite{ji2017adaptive} and wavefront shaping ~\cite{horstmeyer2015guidestar} approaches. 
Adaptive optics (AO) efficiently compensates for optical aberrations. It relies on the weak turbulence assumption and considers that a spatially coherent light beam (e.g. originating from a guide-star or a laser) only experiences smooth wavefront distortions. This assumption remains valid shallow in the sample ($z\lesssim \ell_s$), in transverse planes conjugated with smooth and large-scaled refractive index mismatches. 
For discontinuous, small-scaled or typical volumetric refractive index mismatches, multiple light path trajectories (\emph{i.e.} scattering) are involved and interfere, resulting not only in wavefront distortions but also in highly contrasted intensity fluctuations.
At depths larger than $\ell_s$, the fraction of scattered light exceeds ballistic energy, potentially by several orders of magnitude. More complex optical control is then demanded, referred as ``wavefront shaping'' techniques in the literature~\cite{horstmeyer2015guidestar}.
A system compensating tissue distortion at various depths must be able to measure and correct both low varying wavefront distortions -- attributed to aberrations -- and highly scrambled wavefronts induced by scattering. 

\paragraph{Iterative vs. single-shot phase conjugation.}

Aberrations and scattering corrections rely on the principle of phase conjugation. The latter is based on the use of a spatial light modulator (SLM) placed in the excitation and/or the detection path of an imaging system.
Phase conjugation can then be achieved either iteratively or by direct wavefront measurement. In the iterative approaches, the correcting pattern is either obtained by fully characterizing the transmission matrix from several measurements~\cite{popoff2010measuring,badon2016smart} or by sequentially updating the pattern at the SLM in order to maximize a proper metric that ensures focusing. This metric can be the intensity at a given position~\cite{vellekoop2007focusing}, a non-linear signal~\cite{katz2014noninvasive,may2021fast,cui2010implementation,blochet2021fast}, an image quality metric~\cite{debarre2009image, ji2010adaptive} or a variance contrast~\cite{boniface2019noninvasive}. 
Those approaches have proven successful to correct both scattering and aberration in a wide range of scenarios~\cite{horstmeyer2015guidestar}: coherent ~\cite{vellekoop2007focusing}, incoherent ~\cite{vellekoop2008demixing}, with ~\cite{vellekoop2008demixing} or without guide-stars ~\cite{katz2014noninvasive,rauer2022scattering}.
However, these techniques demand multiple sequential measurements at the expense of speed, which ultimately becomes an issue, especially in the case of rapidly evolving media such as living animals~\cite{qureshi2017vivo}.

Alternatively, direct wavefront conjugation is also possible and potentially faster. It consists in first measuring the wavefront arising from a guide-star (and emerging from the scattering medium) and then to correct wavefront distortions in a second step. This solution is effective when using coherent guide-stars~\cite{liu2015optical,ruan2015optical} since interferometric measurements is then possible. 
Conversely, for incoherent guide-stars such as fluorescent, the outgoing wavefield can only be measured with reference-less wavefront sensing techniques. 
However, in this case, only multi-acquisition schemes combined with phase-retrieval optimization algorithms have allowed measuring complex fluorescence wavefronts through scattering media so far~\cite{aizik2022fluorescent,baek2023generalized}, and single-shot fluorescence wavefield measurement has been limited to low-order aberration compensation in AO schemes~\cite{ji2017adaptive,Booth_NR_21,Fragola_OE_22}. Noteworthy, direct scattering compensation by single-shot amplitude-only acquisition  was also demonstrated based on a common-path speckle-interferometric measurement~\cite{vellekoop2012dopcFluo} but then ignoring ballistic low-order aberrated light. Furthermore, in ref.~\cite{vellekoop2012dopcFluo}, DOPC by binary amplitude-only beam modulation was achieved, at the expense of a drastic loss in focusing efficiency~\cite{Mosk_OE_11,liu2017focusing}.
Therefore, a single-shot measurement of both aberrant and scattered fluorescence wavefronts originating from an incoherent guide-star hidden by a complex medium has never been achieved. 
The solution we propose relies on a typical AO scheme. 

\paragraph{Wavefront sensors.}
Fluorescence has established itself as the gold standard optical contrast for bio-microscopy since being minimally invasive, bio-specific, compatible with living samples, and even having the possibility to be genetically encoded through an increasing number of optogenetic tools, including a large choice of fluorescent proteins with a wide range of photo-sensitive properties. For this reason, fluorescence has also been primarily exploited to perform AO and compensate low-order aberrations~\cite{ji2017adaptive}. 
In bio-imaging, AO is typically implemented by measuring the wavefronts with a Shack-Hartman wavefront sensor (WFS)~\cite{rueckel2006adaptive,tao2011adaptive}. WFS are compact single-shot reference-free instruments, compatible with broadband light sources and providing the complex amplitude of a light-beam (phase and intensity). They are thus ideally suited to measure fluorescence signals.
WFS actually measures the wavefront gradient and thus requires a numerical integration step.
So far, the use of WFS has mostly been limited to smooth and low-order aberrations measurements~\cite{ji2017adaptive,Booth_NR_21,Bon_NM_18} although they have proved to be capable of providing high spatial resolution of smooth objects~\cite{primot1995achromatic,bon2009quadriwave}. 
In the context of astronomy, it was shown that in the case of strong turbulences, singular points of zero intensity associated with screw-phase dislocations (or optical vortices) appear and degrade AO performances~\cite{Tyler_JOSA_00}, especially because deformable mirrors cannot compensate phase dislocations. AO, in its traditional implementation, has thus only been considered for compensating the first few low-order Zernike polynomials contribution to aberrations~\cite{Booth_NR_21} and has long ignored the contribution of optical vortices to the phase structures, so preventing its use for compensating high-order multiple-scattering processes.
Measuring random speckle wavefields has long remained a challenge because of the high spatial density of intrinsic optical vortices~\cite{Baranova_JEPT_81,Berry_PRSLA_74,pascucci2016superresolution}, that are responsible for a so-called ``hidden phase''~\cite{fried1998branch} canceled by the integration step~\cite{Asundi_OLE_15}. Optical vortices are associated with a vanishing intensity at the singularity location and a non-conservative and diverging grid-distortion vector-field. 
Numerical~\cite{LeBigot_OL_99} and experimental~\cite{Dainty_OE_12} efforts have been made to detect optical vortices using Shack-Hartmann WFS, but despite Fried's suggestion to reconstruct the corresponding ``hidden-phase''~\cite{fried1998branch}, only individual isolated vortices could be successfully restored in practical experiments~\cite{LeBigot_OL_98,Starikov_OL_07,Dainty_OE_10}.
Recently, we proposed a solution that experimentally demonstrated the possibility to quantitatively rebuild complex wavefields containing high densities of optical vortices~\cite{wu2021reference}. By performing image processing of the Helmholz decomposition, and relying on the quantization of optical vortex charges, we demonstrated accurate experimental speckle measurement with a reference-less WFS (see Fig.~\ref{fig:comparison_WFS_DH}).

\paragraph{Spectral sensitivity of scattering samples.}
Although WFS are compatible with broadband guide-stars, performing high-resolution digital optical phase conjugation (DOPC) using broadband light beams like fluorescent signals and/or ultrashort laser pulses through multiple scattering samples is complicated. Multiple scattering usually involves large distributions of path length trajectories, and thus results in narrow spectral-correlation bandwidths, and blurred broadband speckle patterns of low contrast whose phase and amplitude are not defined. 
Nevertheless, biological tissues are multiple scattering samples having the specific property to exhibit large anisotropy factors~\cite{jacques2013optical}. They can be typically modeled numerically by stacking thin low-angle-scattering phase plates~\cite{schott2015characterization,cheng2019development,arjmand2021three}. 
We have demonstrated that despite the multiple scattering process, the large anisotropy factor $g$ of biological tissues is responsible for large spectral bandwidths~\cite{vesga2019focusing,zhu2020chromato,arjmand2021three} scaling as $\Delta \lambda = \lambda^2\ell^\ast/L^2$, where $L$ is the slab thickness ($\ell_s \ll L \ll \ell^\ast$) and $\ell^\ast=\ell_s/(1-g)$ is the transport mean-free-path. An ``achromatic'' speckle plane lying in a virtual image plane located at $2/3$ of the slab thickness was identified~\cite{zhu2020chromato,arjmand2021three} (See Supp.~Fig.~\ref{fig:achromatic_plane}). 
The intermediate regime of biological tissues thus bridges the gap between the community of AO and wavefront shaping~\cite{papadopoulos2017scattering,Aubry_SA_20,may2021fast,Mastiani:22,thendiyammal2020model} by involving both the contribution of ballistic light (altered by geometrical aberrations) on the one hand, and the contribution of multiply scattered light on the other hand. 

Here, we demonstrate the possibility to compensate multiple scattering in a single-shot DOPC experiment in the absence of a reference beam. The system we use, based on a high-resolution WFS, is shown to be compatible with broadband light sources like fluorescent guide-stars. This demonstration is based on a rigorous quantitative analysis of spectral correlations of multiply forward scattering samples and the resulting highly multimodal speckle beams. First, we demonstrate the key contribution of optical vortices to the wavefield structure for phase conjugation applications. Second, we discuss the robustness of speckle phase conjugation to spectral shift thanks to the large spectral correlation width of forward multiply scattering samples~\cite{vesga2019focusing,zhu2020chromato,arjmand2021three}. Third, DOPC performances are experimentally investigated as a function of the required number of photons per spatial modes. Finally, based on the former quantitative analysis, 
we demonstrate digital ($30~{\rm nm}$-wide) fluorescence phase conjugation (DFPC) through $500~{\rm \mu m}$ of paraffin samples.

\section*{DOPC with a wavefront sensor}

\begin{figure}[h!]
\centering
\fbox{\includegraphics[width=0.97\linewidth]{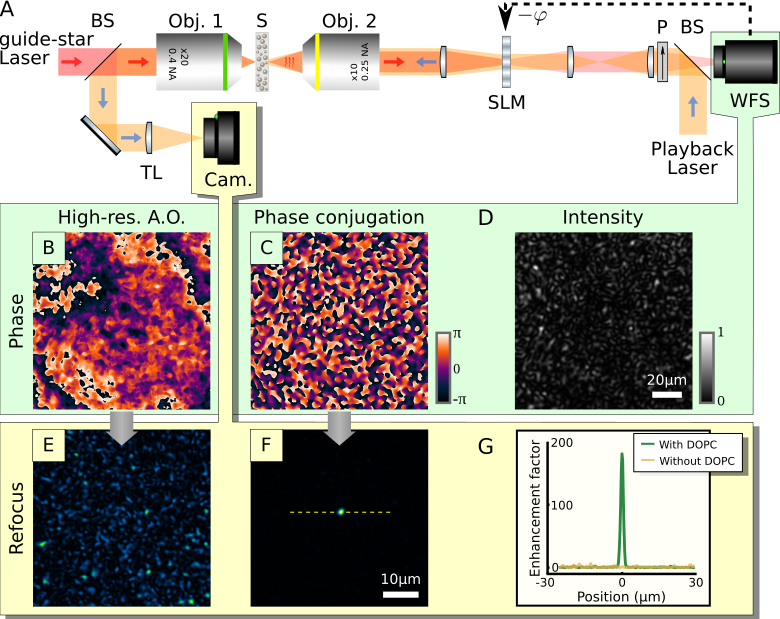}}
\caption{{\bf Principle of DOPC with a WFS.} Scheme of the DOPC experiment (A). A guide-star laser beam is focused by a objective lens (Obj.1) onto the input face of a scattering sample S. The scattered light, collected by a second objective (Obj.2), is sent onto a phase-only spatial light modulator (SLM) conjugated to a WFS (WFS) through a polarizer P. The phase (C) and the intensity (D) of the beam are measured while the SLM is exhibiting a flat phase. The smooth contribution of phase (C) is also rebuilt (B) for comparison. In a second step, a playback laser beam is injected backward thanks to a pellicle beam-splitter. With no phase modulation at the SLM, a speckle pattern (E) is observed on the camera (Cam.). When displaying the measured phase pattern (C) onto the SLM, a sharp focus is observed behind the $720~{\rm \mu m}$-thick spinal chord slice from a mouse (F), exhibiting a $\simeq 200$ enhancement factor (G).}
\label{fig:setup}
\end{figure}

The scheme of the experiment is shown in Fig.~\ref{fig:setup}. A DOPC microscope was built up based on a liquid-crystal SLM (LCoS X10468-01, Hamamatsu, Japan) conjugated with a high-resolution WFS in an ``in-line'' configuration to simplify the alignment of the WFS and SLM~\cite{Mertz_Optica_15,mididoddi2020high} (See Supp.~Fig.~\ref{fig:alignment_SLM_WFS}). In contrast with usual ``pupil'' AO systems, in ``conjugated AO'', both the SLM and the WFS are conjugated to the scattering sample plane~\cite{Mertz_Optica_15}. As a result, the maximum number of spatial modes arriving at the WFS does not depend on the sample but is solely limited by the pupil size of the imaging objective lens (Obj.1 in Fig.~\ref{fig:setup}). For usual ``pupil'' AO designs, the speckle grain size at the SLM/WFS decreases with an increasing size of the illuminated sample region, so resulting in correspondingly large number of modes. This chosen ``conjugate'' AO configuration further allows us to conjugate both the WFS and the SLM to the achromatic plane of the forward scattering samples, which also turns out to be the best conjugating plane for optimizing the isoplanatic patch~\cite{osnabrugge2017generalized}. 

The WFS we use here is a home-made quadri-wave lateral shearing interferometer~\cite{primot1995achromatic} made of a checkerboard phase grating conjugated to a plane located at a few millimeters from a 4Mpx sCMOS camera (Prime BSI Express, Teledyne-Photometrics). The possibility to accurately measure the full speckle phase pattern from a WFS based on the algorithm of Ref.~\cite{wu2021reference} is shown in Supp.~Fig.~\ref{fig:comparison_WFS_DH} by comparing with a digital holographic measurement. In the DOPC setup, the maximum number of spatial modes arriving at the WFS was tuned to $\simeq 3000$ by optical design. This value was chosen based on a prior quantitative experimental characterization of the WFS using the setup shown in Fig.~\ref{fig:setup_WFS_DH}. Also based on the experimental results of this first scaling step, the expected evolution of the DOPC performances is plotted as a function of the number of modes in Fig.~\ref{fig:WFS_DH_number_of_modes}. The process for optimizing the distance between the phase mask grating and the camera is described in Supp.~Fig.~\ref{fig:optimization_distance_d}.

\begin{figure}
\centering
\fbox{\includegraphics[width=0.97\linewidth]{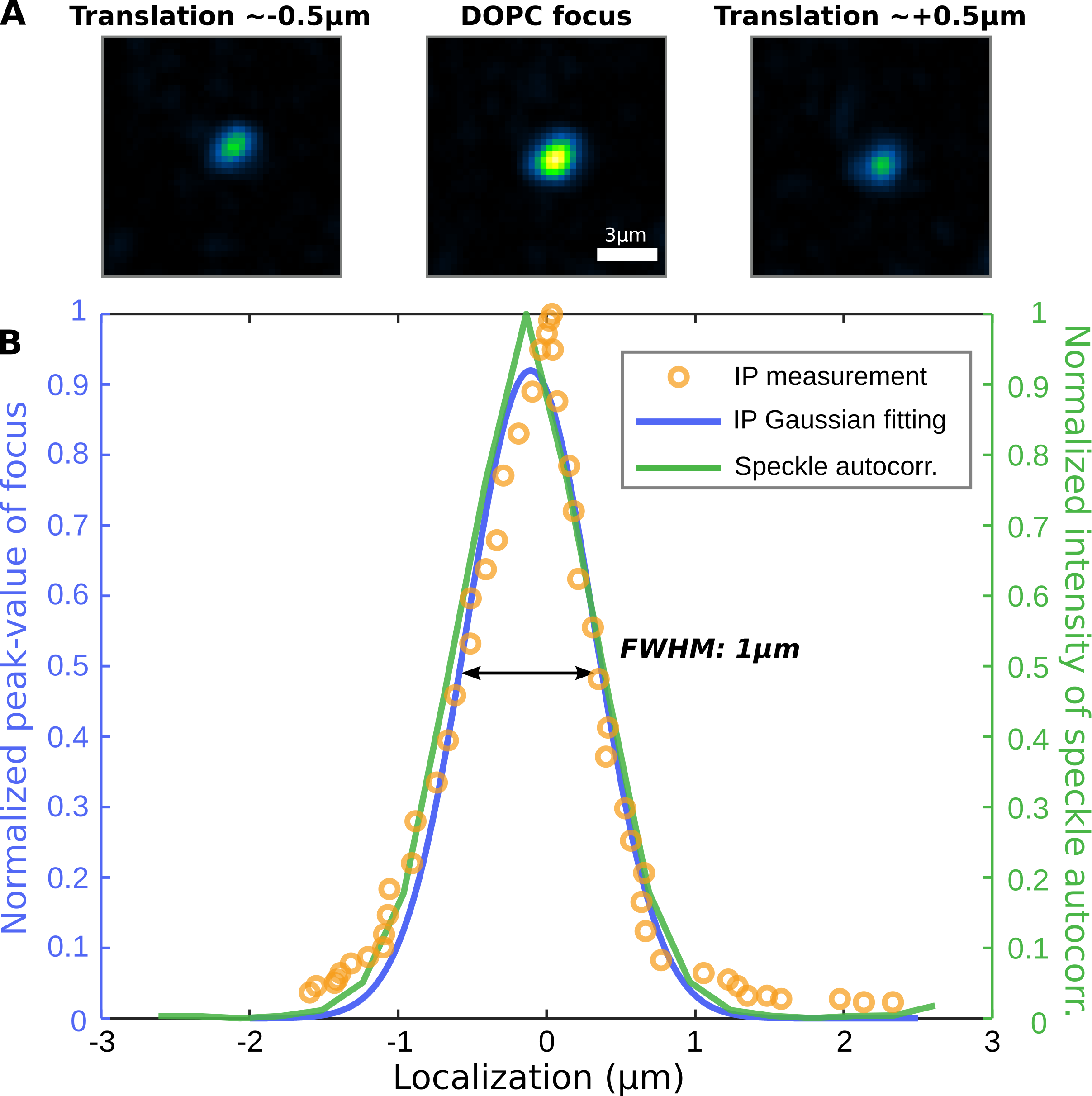}}
\caption{{\bf Size of the isoplanatic patch (IP) through the $720~{\rm \mu m}$-thick spinal cord slice in Fig.~\ref{fig:setup}.} Phase conjugation yields a sharp focus that can be translated by shifting the pattern aside at the SLM (A) and its intensity decreases away from the guide-star location (A). The evolution of the peak intensity of the focus with the amount of translation is plotted in (B) (hollow orange circles) and compared to the autocorrelation plot profile of the speckle pattern measured in the absence of phase conjugation (green curve). The ability to translate the focused spot defines the dimensions of the isoplanatic patch (IP). The matching between the two curves demonstrates that phase conjugation effectively compensates multiple scattering events through the slice.}
\label{fig:IP_spinal_cord}
\end{figure}

First, we demonstrate DOPC with a WFS at a single monochromatic wavelength. In Fig.~\ref{fig:setup}, the principle is illustrated using a fixed $720~{\rm \mu m}$-thick spinal chord slice from a mouse as a sample. In a first step, a guide-star point source is created at the rear of the tissue by focusing a $635~{\rm nm}$ diode laser beam, after prior spatial cleaning through a monomode fiber (see Supp.~Fig.~\ref{fig:setup_DOPC_full}). A flat phase is displayed onto the SLM to record the output wavefield with the WFS, intensity (Fig.~\ref{fig:setup}D) and phase (Fig.~\ref{fig:setup}C). 
In a second step, the phase measured at the WFS (Fig.~\ref{fig:setup}C) is conjugated and addressed to the SLM, which is then illuminated by a back-propagating playback laser at the same wavelength. A sharp focus is then observed at the readout camera (Fig.~\ref{fig:setup}F) exhibiting an enhancement factor~\cite{vellekoop2007focusing,cui2010implementation} as large as $I_{focus}/\left<I_{speckle}\right>\simeq 200$ (Fig.~\ref{fig:setup}G). 
This value can be compared to the maximum theoretical expectation which is qualitatively given by the number of measured and controlled modes~\cite{vellekoop2007focusing,jang2017optical}. 
In our system, this number was set to $\simeq 3000$ by optical design (a factor $\simeq 15$ above the experimental result and a factor $\simeq 10$ above one might have been expecting from prior calibration experiment (Fig.~\ref{fig:WFS_DH_number_of_modes})). The observed difference between the experimental enhancement factor and its theoretical maximum value may involve partial depolarization of the playback laser by the scattering medium, slight misalignment between the SLM and the WFS (Fig.~\ref{fig:alignment_SLM_WFS}), and imperfect wavefront reconstruction. Interestingly, when ignoring the contribution of optical vortices in the phase reconstruction process (Fig.~\ref{fig:setup}B), such as currently done by all Shack-Hartmann-based AO systems, no focus is obtained (Fig.~\ref{fig:setup}E). The reason is that the fraction of ballistic light through this sample is negligible. This result demonstrates that high-resolution WFS enables reference-less single-shot DOPC through multiply scattering media. The ability to compensate multiple scattering is substantiated by the size of the isoplanatic patch, measured to be as small as a single speckle grain size (Supp.~Fig.~\ref{fig:IP_spinal_cord}).
DOPC through a multimode fiber, of interest for micro-endoscopy techniques, is also demonstrated in Supp.~Fig.~\ref{fig:DOPC_fiber}.
The dimensions of the focal spot in Fig.~\ref{fig:setup}F were measured to be $1.2~{\rm \mu m}$-FWHM and $1.5~{\rm \mu m}$-FWHM along two orthogonal directions, respectively. These values are very close to the diffraction limit achievable with our objective lens (Obj.~2 in Fig.~\ref{fig:setup}A): $0.51\lambda/{\rm NA}=1.31~{\rm \mu m}$, where NA is the numerical aperture of Obj.~2. The possibility to get focal spot dimensions smaller than the NA of the focusing lens is made possible thanks to the scattering medium increasing the effective focusing aperture~\cite{vellekoop2007focusing}.
In Supp.~Fig.~\ref{fig:hologram_delivery}, we further demonstrate that using computer generated holography, it is possible to exploit the angular memory effect of a diffuser to project more complex intensity patterns through the scattering sample.

\section*{Contribution of optical vortices to DOPC}

\begin{figure}[h]
\centering
\fbox{\includegraphics[width=0.97\linewidth]{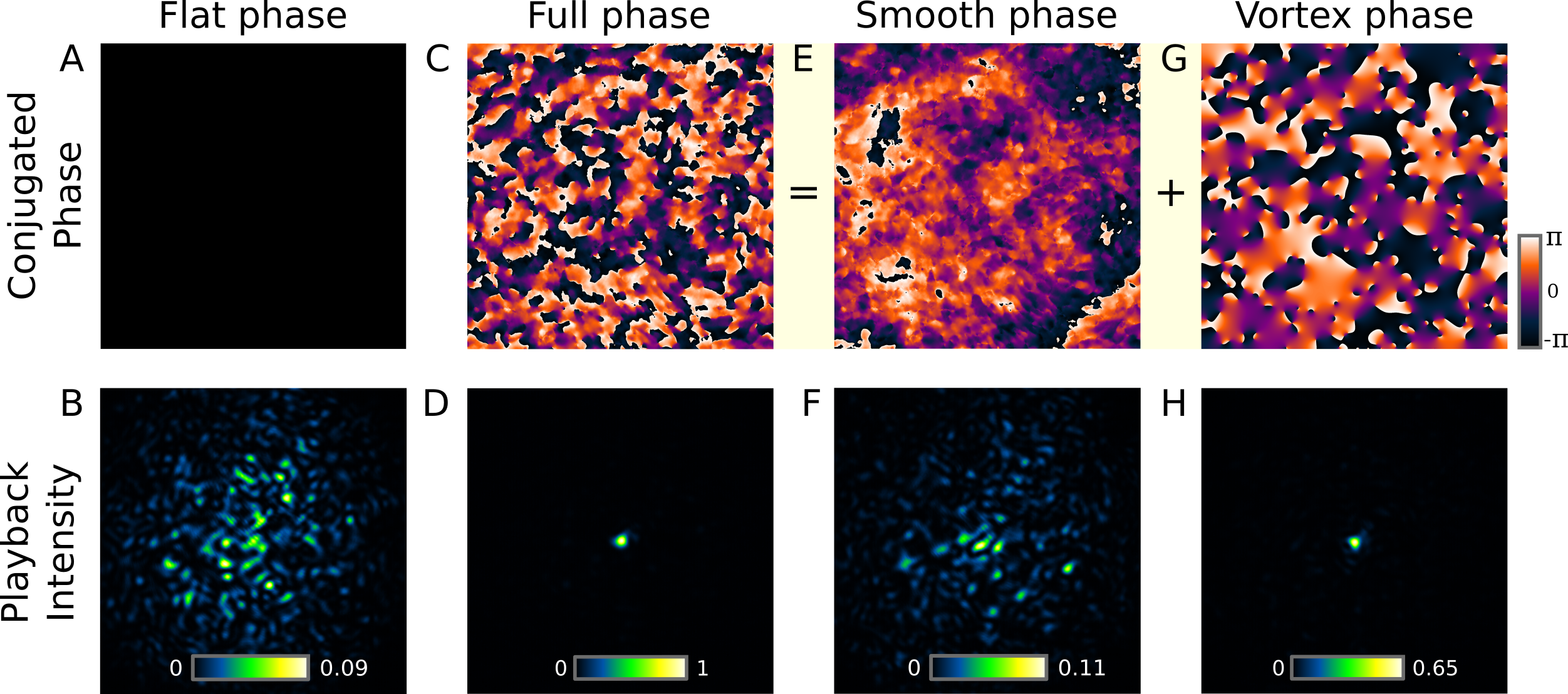}}
\caption{{\bf Importance of the optical vortices for efficient DOPC through an out-of-focus $1^\circ$~surface diffuser.} The SLM and WFS are conjugated to the virtual image of the guide-star focus. Displaying a flat phase on the SLM (A) results in a speckle at the playback camera (B) while full phase compensation (C) yields a focus (D) containing $25\%$ of light energy (amplitude set to one unit). The smooth contribution to the phase (E), typically rebuilt in AO experiments does not provide a focus (F) while the vortex contribution to the phase (G) does (H).}
\label{fig:vortex_or_not_vortex}
\end{figure}

With a Shack-Hartmann-like WFS, the wavefront is obtained by integrating the measured gradient vector field. Usual reconstruction algorithms only considered the curl-free contribution of the vector field, so resulting in a conservative potential~\cite{Bon_AO_12,Asundi_OLE_15}. The non-conservative solenoidal contribution of optical vortices, responsible for the so-called ``hidden phase''~\cite{fried1998branch} has so far been ignored in typical AO schemes. Here, we thus specifically study the contribution of both the conservative phase and the hidden-phase to DOPC performances. Results are shown in Fig.~\ref{fig:vortex_or_not_vortex}. A speckle is created by illuminating a holographic diffuser ($1^\circ$ diffuser, Edmund Optic) with a collimated laser beam. The diffuser was defocused by $2~{\rm mm}$ from the conjugate plane of the SLM/WFS so yielding a speckle pattern containing many optical vortices. When a flat phase is displayed onto the SLM, the playback laser beam yields a fully-developed speckle pattern on the camera (Fig.~\ref{fig:vortex_or_not_vortex}A and~\ref{fig:vortex_or_not_vortex}B). When conjugating the full phase measured at the WFS including both phase contributions, a focus is obtained containing $25\%$ or light energy (Fig.~\ref{fig:vortex_or_not_vortex}C and~\ref{fig:vortex_or_not_vortex}D). Its amplitude was normalized for comparison with other cases. Interestingly, when considering only the smooth (\emph{i.e.} conservative) contribution of the phase, which is solely rebuilt in AO systems, the laser energy is slightly concentrated to the central region but no focus is obtained in this scattering condition which cannot be considered as low-order optical abberations (Fig.~\ref{fig:vortex_or_not_vortex}E and~\ref{fig:vortex_or_not_vortex}F). In contrast, we observed that the vortex phase allows focusing $\simeq 2/3$ of the maximum energy, corroborating the idea that the structure of phase in speckle patterns is driven by vortex phase singularities~\cite{Freund_1001_correlations}. 
Experimentally, the fraction of focused energy was calculated by computing the ratio between the energy in the focused spot and the integration of the whole transmitted speckle signal obtained without phase conjugation, which fully falls in the field of view of our camera for forward scattering media (see Supp.~Fig.~\ref{fig:DOPC_performance_estimation}).
Hereafter, we characterize DOPC performances by measuring the fraction of refocused energy rather than the enhancement factor. The enhancement factor is more suitable to cases where the number of controlled modes is much smaller than the number of modes of the scattering medium, which is the case of the sample used in Fig.~\ref{fig:setup}. Conversely, for more forward scattering samples, the total transmitted energy is measurable and the fraction of refocused energy becomes comparable to one.
This result demonstrates that rebuilding the non-conservative contribution of the gradient vector field measured by WFS is of critical importance to efficiently compensate scattering in DOPC experiments.

\section*{DOPC from a spectrally-detuned guide-star}

\begin{figure}[h]
\centering
\fbox{\includegraphics[width=0.97\linewidth]{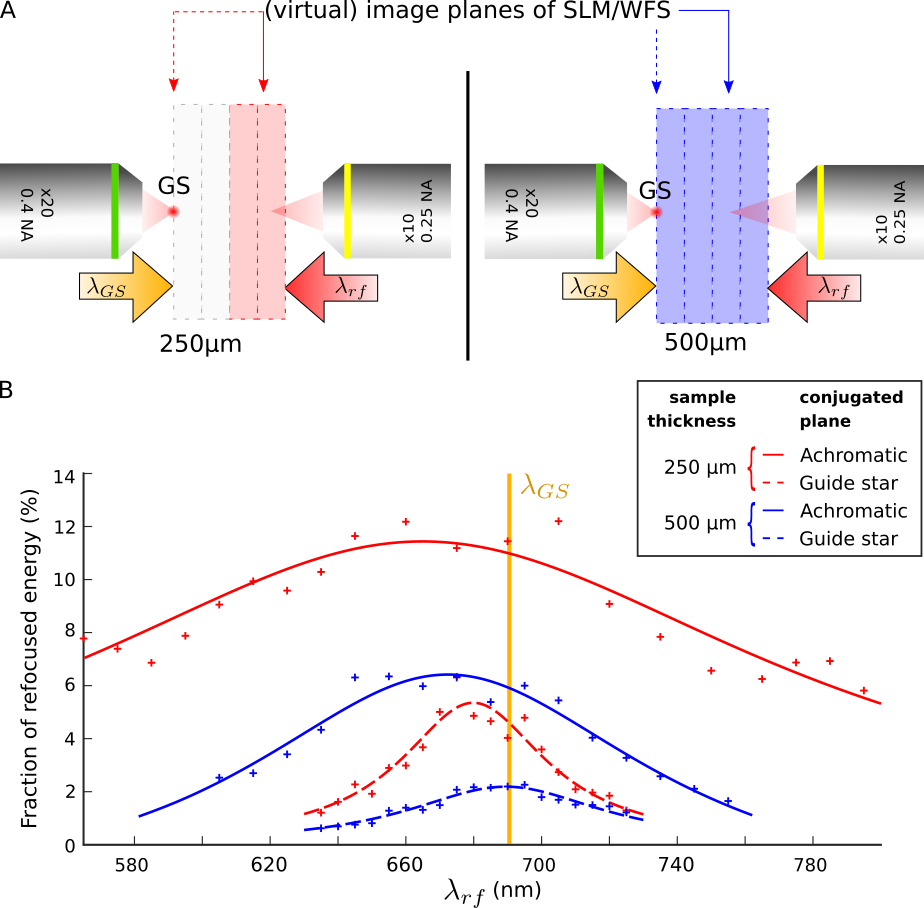}}
\caption{{\bf Fraction of refocused energy as a function of spectral detuning between the refocusing laser $\lambda_{rf}$ and the guide-star laser ($\lambda_{GS} = 690/5nm$).} Samples consist in $250~{\rm \mu m}$-thick (red curves) and $500~{\rm \mu m}$-thick (blue curves) samples of paraffin. In both cases, refocusing efficiency is plotted in two conditions: with the SLM/WFS conjugated to the ``achromatic'' plane (solid arrows in (A) and solid fitting lines in (B)) and with the guide-star plane (dashed arrows in (A) and dashed fitting lines in (B)). 
}
\label{fig:detuned_GS}
\end{figure}

In constrast to monochromatic DOPC that we have just demonstrated, the implementation of DOPC with a fluorescent guide-star faces the specific challenge of robustness to spectral width and detuning. It is indeed necessary to be able to both measure broadband speckles and refocus a laser beam at a wavelength shorter than the average fluorescence wavelength to take into account the $\simeq 25~{\rm nm}$ Stokes' shift of typical organic fluorophores. 
Then, we considered studying the robustness of DOPC to spectral detuning by shifting the wavelength of the playback laser as compared to the guide-star's.
In a former article, we showed that a fixed $1~{\rm mm}$-thick brain slice can exhibit a spectral correlation width as large as $\simeq 200~{\rm nm}$~\cite{vesga2019focusing}. Although experiencing multiple scattering events over optical trajectories much larger than the wavelength, we identified that, in virtue of the large anisotropy factor of such media ($g$ close to $1$), all trajectories are snake-like~\cite{French_JPD_03} and the dispersion in trajectories path lengths is much smaller than their average length~\cite{zhu2020chromato}. Consequently, the spectral bandwidth, scaling as the inverse of the dispersion in optical path delays, was derived to be $\Delta \lambda = \frac{\lambda^2\ell^\ast}{L^2}$. 
Both analytical derivation~\cite{zhu2020chromato} and numerical simulations~\cite{arjmand2021three}, supported by experimental results demonstrated that the speckles generated through the scattering sample had a spectral-correlation maximum in a virtual plane located at $2/3$ of the slab thickness. 
Noteworthy, not only speckle intensities are achromatic in this ``achromatic'' plane but also their spatial phases~\cite{zhu2020chromato} and even their spectral phases~\cite{arjmand2021three}. 
 
\begin{figure}
\centering
\fbox{\includegraphics[height=0.93\linewidth]{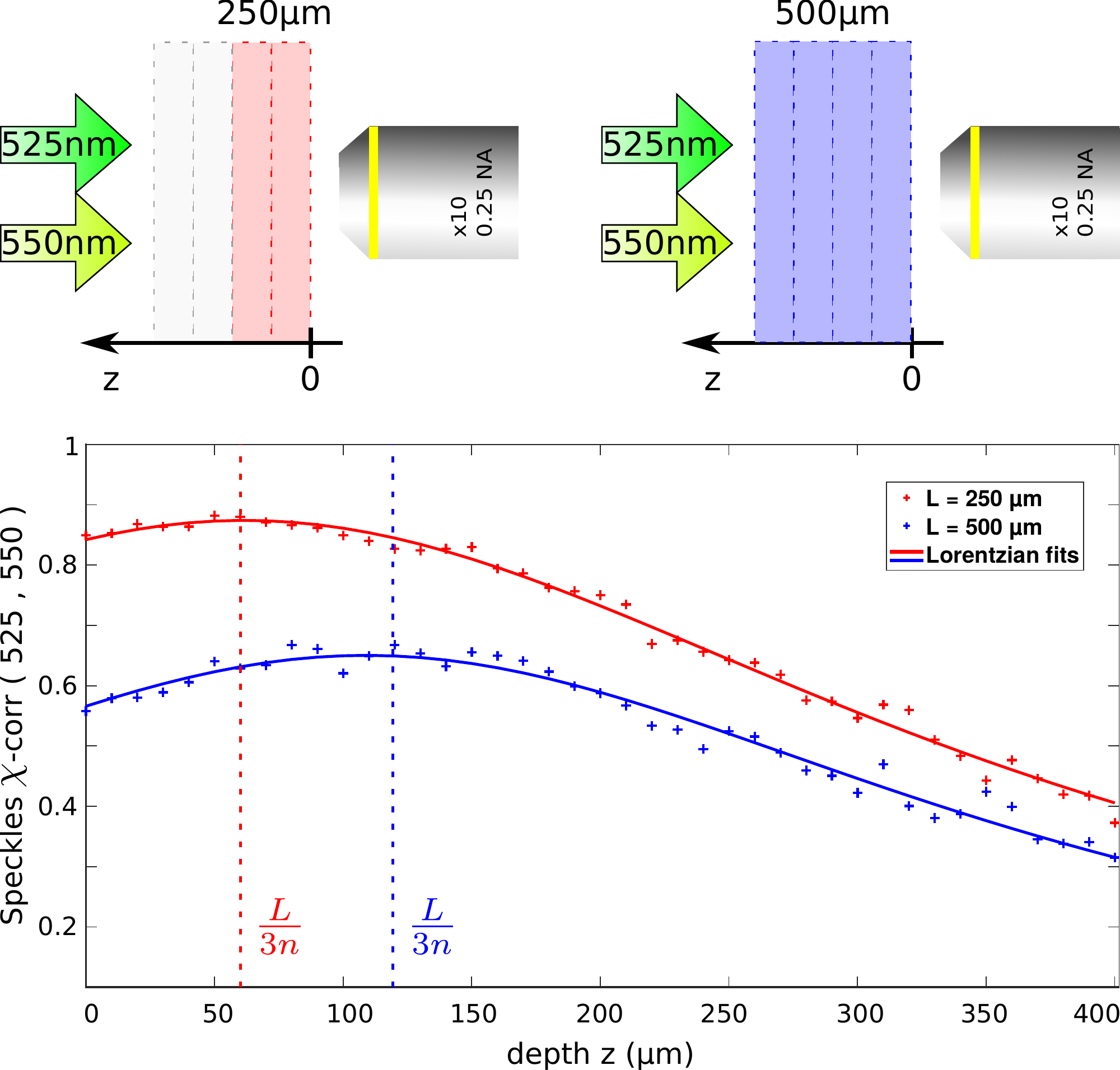}}
\caption{{\bf Achromatic plane location evidenced by axial cross-correlation product of speckles at two nearby wavelengths ($\lambda_1=525~{\rm nm}$ and $\lambda_2=550~{\rm nm}$).} The beams are transmitted through paraffin samples of thicknesses $L=250~{\rm \mu m}$ (red) and $L=500~{\rm \mu m}$ (blue) when illuminated by a collimated laser beam. Experimental data are fitted using Lorentizan curves ($C = 1/[1+(z-z_{achr})^2/w_z^2]$), both revealing a correlation maximum at $z_{achr} \simeq \frac{L}{3n}$, with $L$ the slab thickness and $n$ the average refractive index of paraffin, in agreement with theory. Noteworthy, the $L/3$ position of the achromatic plane must be corrected by the average refractive index mismatch between the slab (paraffin) and the embedding medium (air). Measured achromatic planes: $z_{achr}(250~{\rm \mu m})=61.4~{\rm \mu m}$, $z_{achr}(500~{\rm \mu m})=109~{\rm \mu m}$ \emph{vs.} theoretical achromatic planes: $z_{achr}^{th}(250~{\rm \mu m})=59.5~{\rm \mu m}$, $z_{achr}^{th}(500~{\rm \mu m})=119~{\rm \mu m}$ for an estimated value $n=1.4$.}
\label{fig:achromatic_plane}
\end{figure}

We thus conjugated our SLM/WFS to this ``achromatic'' plane in order to both maximize coherence of the broadband guide-star speckle and to make DOPC robust to spectral detuning. To demonstrate the relevance of this configuration, we compare its performances to the case where the SLM/WFS are conjugated to the guide-star plane. Data shown in Fig.~\ref{fig:detuned_GS} are obtained for two sample thicknesses, $L=250~{\rm \mu m}$ and $L=500~{\rm \mu m}$, obtained by stacking two and four layers of parafilm, respectively.
The achromatic plane location, that we evidenced experimentally (see Supp.~Fig.~\ref{fig:achromatic_plane}), is identified by solid-lined arrows in Fig.~\ref{fig:detuned_GS}A and the guide-star planes by dashed-lined arrows. A focused laser spot having a fixed wavelength $\lambda_{GS} = 690~{\rm nm}$ was used as a guide-star and DOPC performances measured as the playback laser was spectrally scanned ($\lambda_{rf}$). Experimentally measured fraction of refocused energy are plotted in Fig.~\ref{fig:detuned_GS}B and fitted by Lorentzian profiles (in agreement with theory~\cite{zhu2020chromato}):
\begin{equation}
F(\lambda_{rf}) = \frac{F_0}{ 1 + \left(\frac{\lambda_{rf}-\lambda_0}{\Delta\lambda} \right)^2 }
\end{equation}
Fitting parameters are given in Table~\ref{table:spectral_width}.

\begin{table}[h!]
\begin{center}
\begin{tabular}{|l|c|c|c|}
\hline
& $F_0$ ($\%$) & $\lambda_0$ (nm) & $\Delta\lambda$ (nm)  \\
\hline
\multicolumn{4}{|c|}{SLM/WFS conjugated to the {\bf achromatic plane}}\\
\hline
$L= 250~{\rm \mu m}$ & $11.4$ & $665$ & $126$ \\ 
$L= 500~{\rm \mu m}$ & $6.5$ & $671$ & $53.7$ \\ 
\hline
\multicolumn{4}{|c|}{SLM/WFS conjugated to the {\bf guide-star plane}} \\
\hline
$L= 250~{\rm \mu m}$ & $5.34$ & $680$ & $26.1$ \\ 
$L= 500~{\rm \mu m}$ & $2.23$ & $689$ & $36.0$ \\ 
\hline
\end{tabular}
\end{center}
\caption{\label{table:spectral_width} Lorentzian fitting parameters for DOPC efficiency as a function of spectral detuning.}
\end{table}

\begin{figure}
\centering
\fbox{\includegraphics[height=0.97\linewidth]{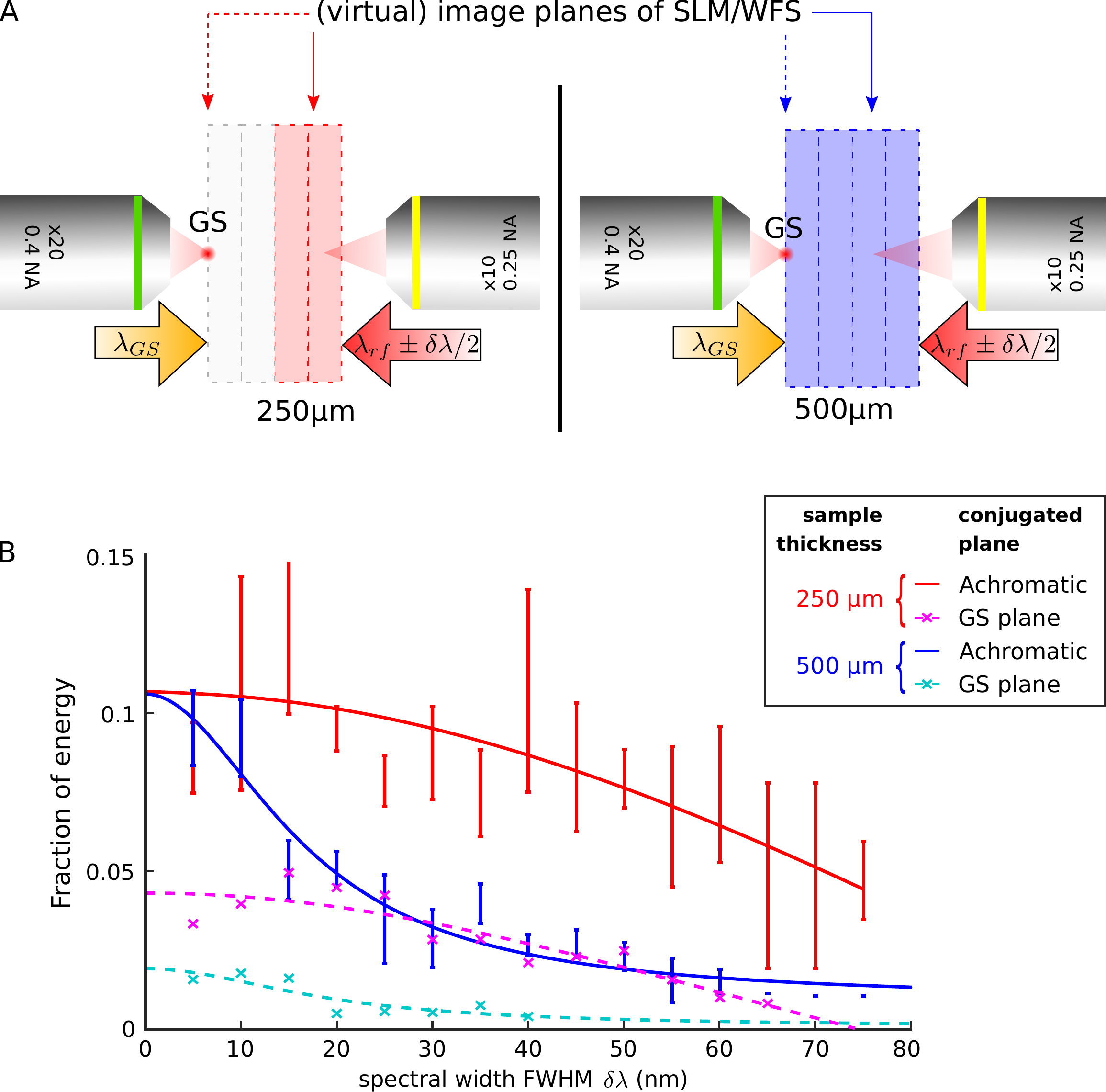}}
\caption{{\bf Effect of the spectral width of the playback laser on the fraction of refocused light energy.} Both for the cases of paraffin samples of thicknesses $L=250~{\rm \mu m}$ (red) and $L=500~{\rm \mu m}$ (blue), the guide-star ($\lambda_{GS}=690~{\rm nm}$) was focused at a physical distance of $500~{\rm \mu m}$ from the output surface of the sample as shown in (A). DOPC was then achieved using a broadband playback light beam of central wavelength $\lambda_{rf}=\lambda_{GS}$ and of bandwidth $\delta\lambda$ FWHM. Fraction of refocused energy as a function of $\delta\lambda$ is shown in (B) for the SLM/WFS conjugated to the achromatic plane (solid line fits) and the guide-star plane (dashed line fits). Experimental data are fitted using Lorentizan functions $f(\delta\lambda) = \alpha/[1+(\delta\lambda/\Delta\lambda)^2]$, with $\Delta\lambda_{achr}(L=250)=70.4~{\rm nm}$, $\Delta\lambda_{GS}(L=250)=44~{\rm nm}$, $\Delta\lambda_{achr}(L=500)=22~{\rm nm}$, $\Delta\lambda_{GS}(L=500)=19.7~{\rm nm}$. For the measurements in the achromatic plane, a set of three realizations was achieved and the experimental variance is shown as bars. }
\label{fig:spectral_width}
\end{figure}

As expected, a larger spectral width is observed when conjugating the SLM/WFS to the achromatic plane for both samples, and the largest spectral width and peak energy are obtained for the thinnest sample. The maximum fraction of refocused energy is $\simeq 12\%$ for the $L=250~{\rm \mu m}$-thick sample and roughly half for $L=500~{\rm \mu m}$. Less expected is the fact that a larger focused energy is obtained at the achromatic plane than at the guide-star plane, even at the guide-star wavelength.
We partially attribute the improvement of DOPC at the achromatic plane to the smaller density of optical vortices in this plane -- roughly half the one in the guide-star plane -- so potentially resulting in less phase reconstruction errors. 
Surprisingly also, the peak of the focused-energy distribution is slightly blue-shifted with respect to the guide-star wavelength. This shift is in favor of one photon fluorescence excitation but its explanation is not obvious (we ruled out a possible SLM-calibration error), all the more because it differs between the cases where the SLM/WFS are conjugated to the achromatic plane ($\delta\lambda \simeq 22~{\rm nm}$) or the guide-star plane ($\delta\lambda \simeq 5~{\rm nm}$).
Importantly, we experimentally checked that the DOPC spectral widths measured by detuning the playback laser line are consistent with both the speckle correlation width of the medium (Supp.~Fig.~\ref{fig:correlation_width}) and with the DOPC spectral widths measured by broadening the spectrum of the playback laser (Supp.~Fig.~\ref{fig:spectral_width}). Our results are thus relevant to non-linear optical imaging modalities requiring multi-line or broadband ultra-short laser beams.

\section*{Required photon budget}

\begin{figure}[h!]
\centering
\fbox{\includegraphics[width=0.97\linewidth]{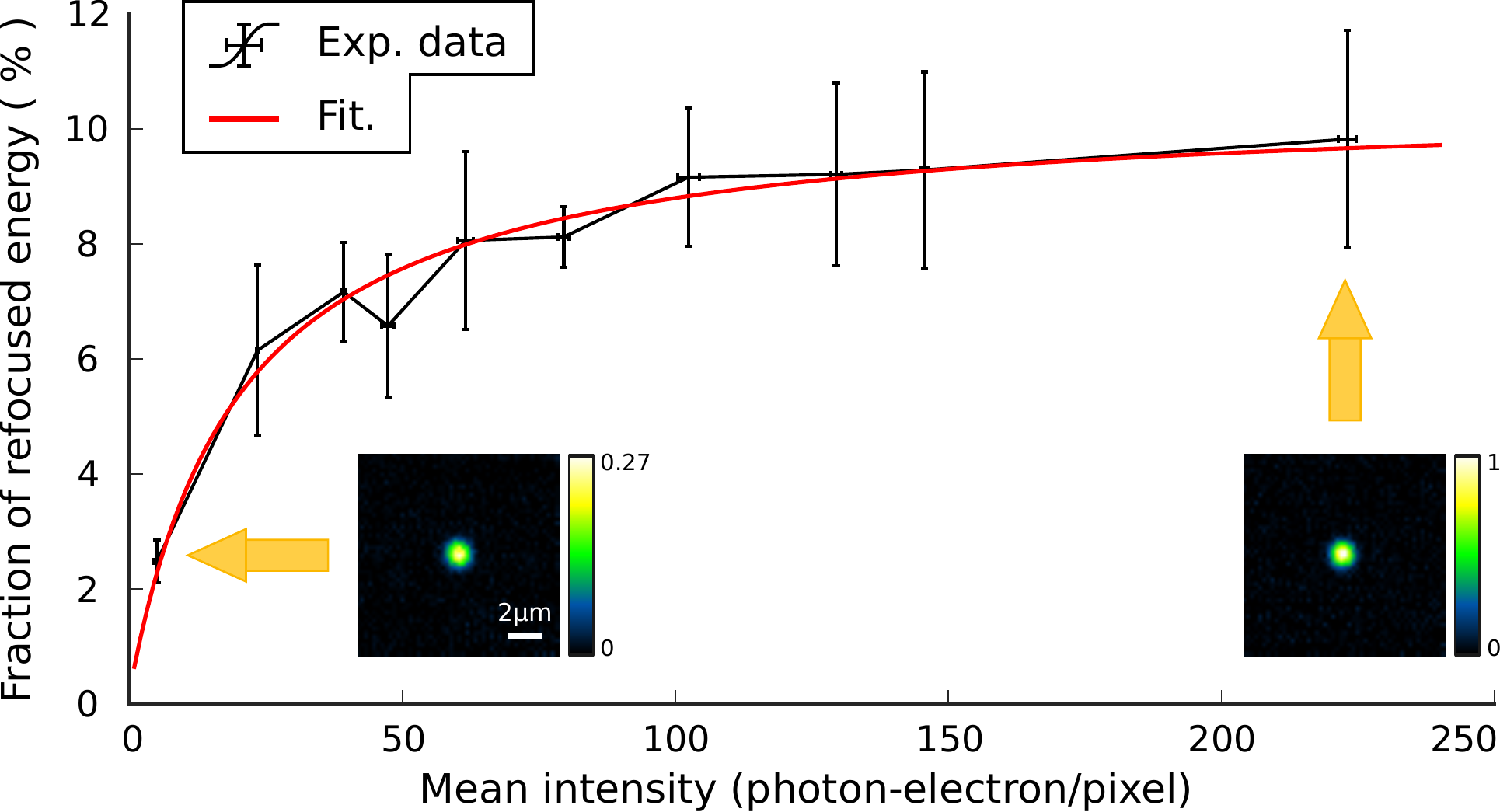}}
\caption{{\bf Efficiency of DOPC as a function of the guide-star photon budget.} The fraction of refocused energy is measured as a function of the average number of collected photon per camera pixels at the WFS and fitted. Here, the speckle patterns were generated using a $1^\circ$ holographic diffuser located 3mm away from the focused laser guide-star at $635~{\rm nm}$. The SLM and WFS are conjugated to the virtual image of the guide-star (like in Fig.~\ref{fig:vortex_or_not_vortex}). Vertical bars represent twice the experimental standard deviations ($\pm \sigma$) computed over 5 measurements.
}
\label{fig:photon_budget}
\end{figure}

The photon budget available from guide-stars for performing DOPC is usually low either because of speed requirement or just because of the limited photon budget of the guide-star itself. Furthermore, through a scattering medium, this energy is split among all outgoing spatial modes. Under these conditions, increasing the numerical aperture of the collecting objective lens ($.25~{\rm NA}$ in our case) does not help since it does not increase the spatial-mode density of photon at the diffuser output, which solely matters.
We thus investigated the photon budget required per spatial mode to perform WFS-based speckle phase conjugation. In Fig.~\ref{fig:photon_budget}, the fraction of focused energy $F$ is plotted as a function of the mean speckle intensity at the WFS $\left<I\right>$, expressed in photo-electron per WFS-pixel. The measurements are fitted with a theoretical model from Ref.~\cite{jang2017optical}: $F = F_0\left(1+\frac{I_0}{\left<I\right>}\right)^{-1}$, with $I_0$, the required average intensity to get half the optimal performances. 
Experimentally, we obtain that this model allows a good fitting of experimental data for a parameter $I_0\simeq 20$ photo-electrons per pixel. 
Considering that our DOPC system was designed to measure and control $\simeq 3000$ spatial modes with our $4$~Mpx WFS-camera, our measurements yield a required photon flux of $\simeq 3\times10^4$ photo-electons per spatial mode. We could also measure that the ratio between the playback speckle extent (without DOPC) and the playback focal spot size (with DOPC) is $\simeq 62$, so giving a number of spatial modes for the diffuser equal to $\pi/4*62^2=3019$, in excellent agreement with our optical design estimate.

\section*{Single-shot DFPC from a fluorescent guide-star}

\begin{figure*}[h!]
\centering
\fbox{\includegraphics[width=0.985\linewidth]{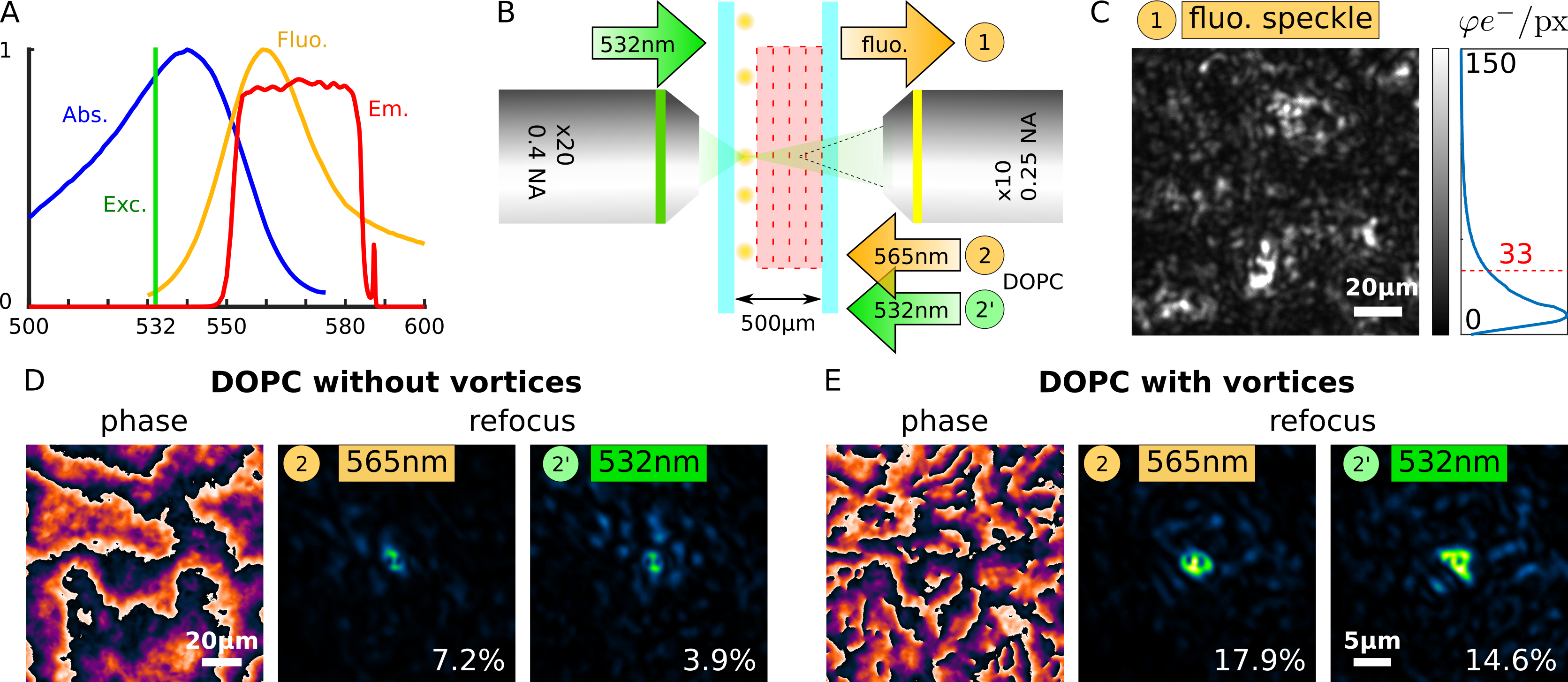}}
\caption{{\bf DFPC using $1~{\rm \mu m}$ fluorescent beads as guide-stars.} The absorption (Abs.) and fluorescence emission (Fluo.) spectra of the beads are shown in (A) together with the excitation laser line at $532~{\rm nm}$ (Exc.) and the band-pass emission filter $565/30$ (Em.). 
Beads are hidden behind a thickness $L=500~{\rm \mu m}$ of a paraffin samples (B). The SLM/WFS are conjugated to the achromatic plane of the scattering samples and the measured fluorescence speckle is shown in (C) together with its intensity histogram exhibiting a mean fluorescence intensity of $33$ photo-electrons per pixel. The playback laser beams is then sent through the sample after phase conjugation both at the average emission wavelength ($\lambda = 565~{\rm nm}$) and at the excitation wavelength ($\lambda = 532~{\rm nm}$) when considering the smooth phase contribution only (D) or the full reconstructed phase (E). The focii quality in d and e are degraded by the refractive index mismatch between the fluorescent micro-bead and the embedding PVA matrix.}
\label{fig:fluo}
\end{figure*}

\begin{figure}
\centering
\fbox{\includegraphics[width=0.97\linewidth]{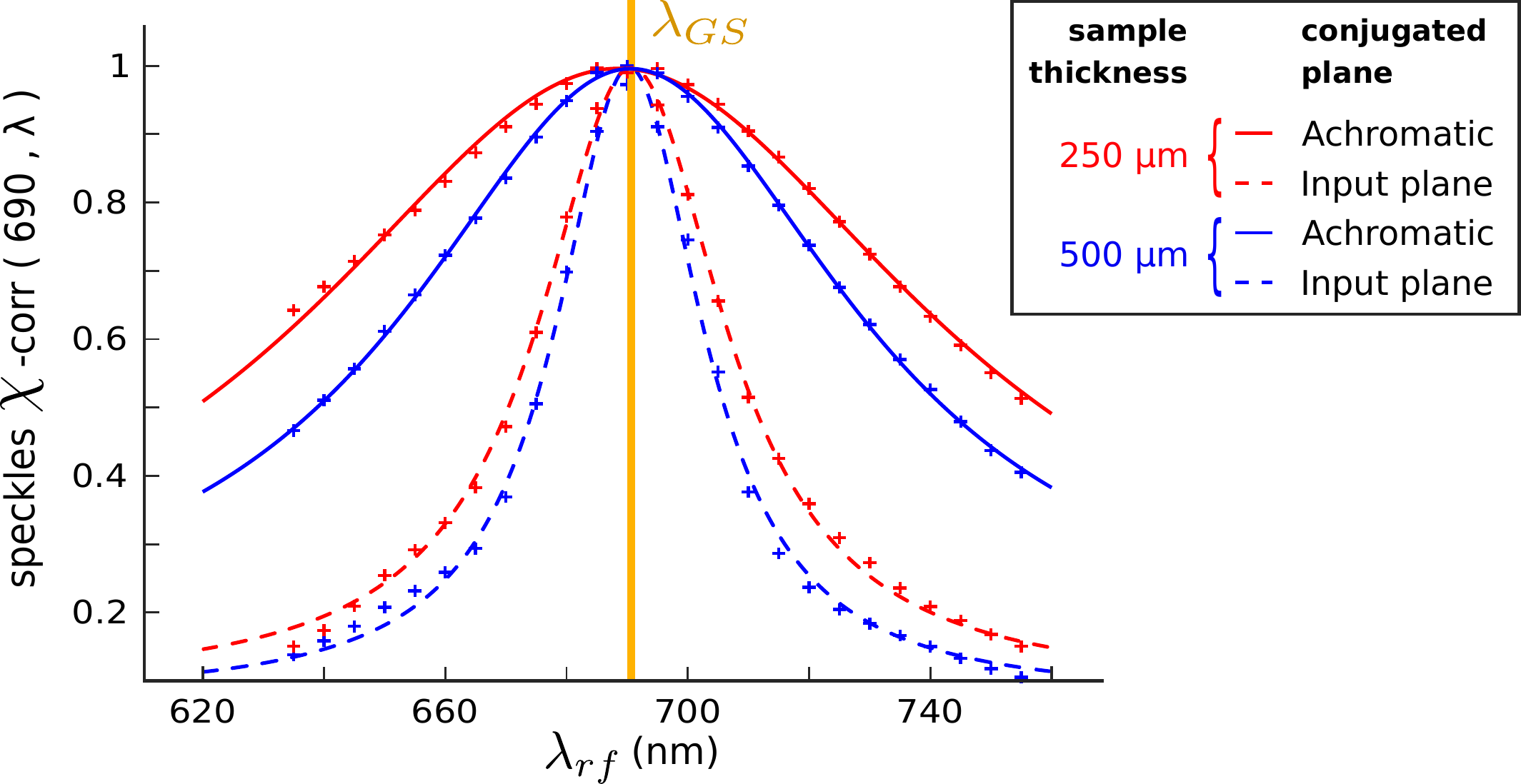}}
\caption{{\bf Spectral cross-correlation product of speckles transmitted through paraffin samples of thicknesses $L=250~{\rm \mu m}$ (red) and $L=500~{\rm \mu m}$ (blue) when illuminated by a collimated laser beam.} A reference speckle is considered at the arbitrary wavelength $\lambda_{GS}=690~{\rm nm}$. Experimental data are fitted using Lorentizan curves ($C(\lambda_{rf}) = 1/[1+(\lambda_{GS}-\lambda_{rf})^2/w_\lambda^2]$) both in the case when the camera records speckles while conjugated to the virtual entrance plane image (dashed lines) and the virtual achromatic plane image (continuous lines). For $L=250~{\rm \mu m}$ (red plots), the measured FWHM are $2w_\lambda({\rm GS}) = 37~{\rm nm}$ and $2w_\lambda({\rm achr}) = 126~{\rm nm}$ in the entrance plane and the achromatic plane, respectively. For $L=500~{\rm \mu m}$ (blue plots), the measured FWHM are $2w_\lambda({\rm GS}) = 29~{\rm nm}$ and $2w_\lambda({\rm achr}) = 90~{\rm nm}$ in the entrance plane and the achromatic plane, respectively. }
\label{fig:correlation_width}
\end{figure}

\begin{figure*}
\centering
\fbox{\includegraphics[height=0.35\textheight]{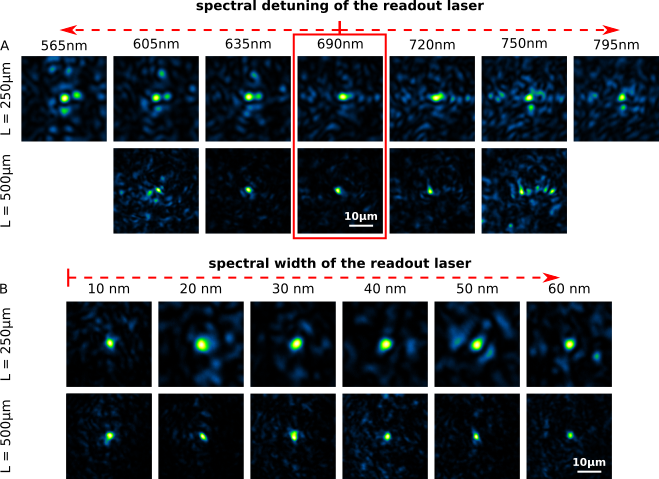}}
\caption{ {\bf Images of the focii of the readout laser.} Two paraffin samples of thicknesses $L = 250~{\rm \mu m}$ and $L = 500~{\rm \mu m}$ are considered. The playback laser has wavelength detuned from the guide-star laser (A) and spectral width increased (B). In B, the spectrum is centered on the guide-star wavelength. Images in A and B correspond to data-points of curves shown in Fig.~\ref{fig:detuned_GS} and Supp.~Fig.~\ref{fig:spectral_width}, respectively. }
\label{fig:imagettes}
\end{figure*}

\begin{figure*}
\centering
\fbox{\includegraphics[height=0.35\textheight]{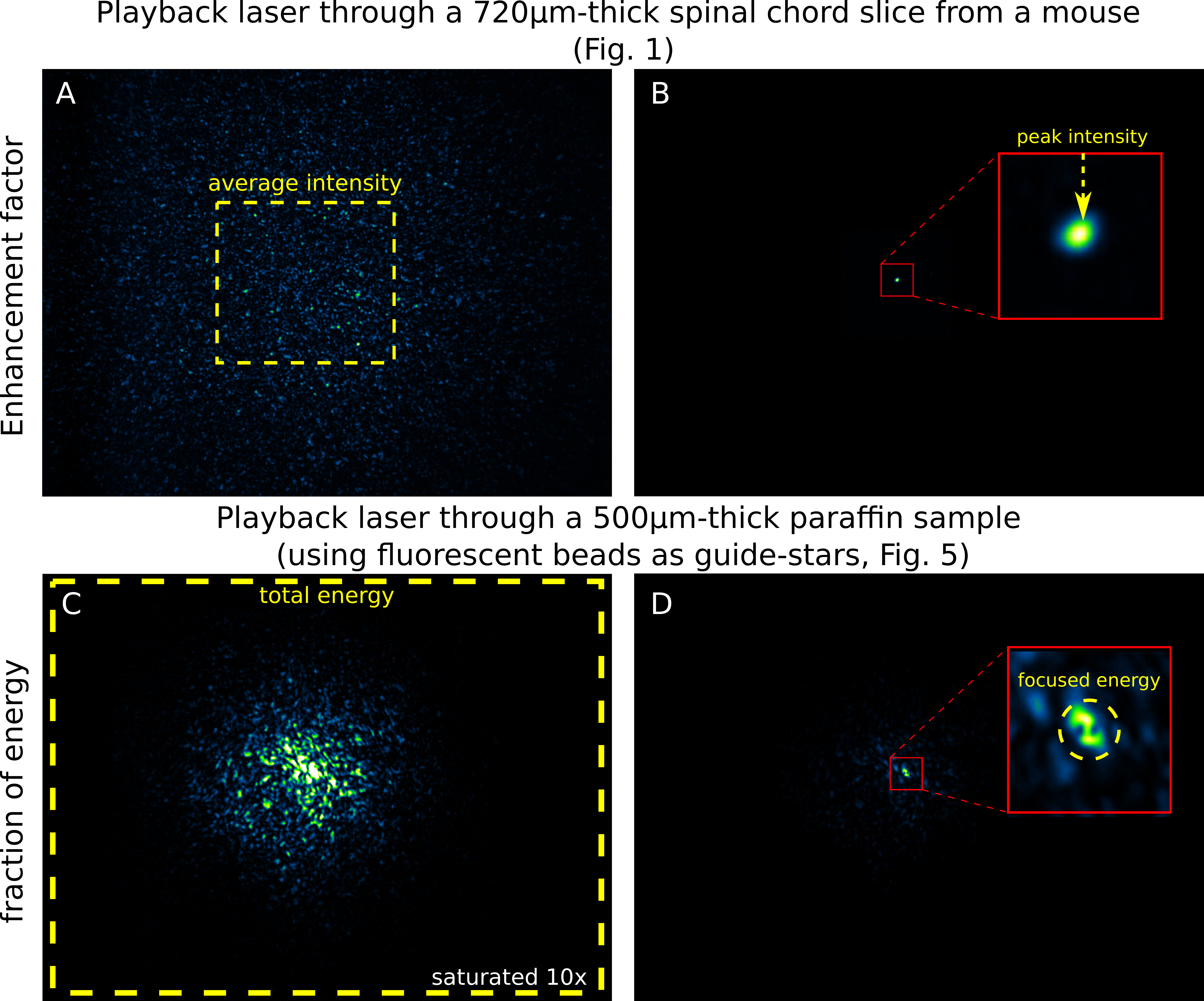}}
\caption{{\bf Experimental estimation of the enhancement factor and the fraction of refocused energy, depending on the scattering sample at play.} For the experiment with the spinal chord slice in Fig.~\ref{fig:setup}, the energy is not fully collected by the objective lens. The enhancement factor is thus considered, defined as the ratio between the peak intensity of the focus (B) and the average speckle intensity (computed in the yellow rectangle where the speckle is uniform in (A)). Conversely, for the DFPC experiment shown in Fig.~\ref{fig:fluo}, light is scattered at smaller angles allowing full light-energy collection by the objective lenses and not allowing measuring the peak intensity of the focus because of the focus degradation by the microbead. The fraction of refocused energy can then be measured, and is then computed as the ratio between the energy in a small circle surrounding the bead (D) and the total light energy on the camera (integrating the energy in the yellow rectangle in (C)). In both conditions, the background noise was subtracted.}
\label{fig:DOPC_performance_estimation}
\end{figure*}

Finally, we demonstrate the ability of our system to perform single-shot DFPC using fluorescent micron-size spheres. Fluorescence excitation was achieved using a $532~{\rm nm}$ laser line and fluorescence was collected by stacking a pair of two emission filters to efficiently block the excitation light (Fig.~\ref{fig:fluo}A). The resulting filter bandwidth of $30~{\rm nm}$ has been chosen so as to match as much as possible the spectral width of fluorescence, within the limit of the spectral correlation bandwidth of the scattering sample (Fig.~\ref{fig:detuned_GS}B and Supp.~Fig.~\ref{fig:correlation_width}). The beads were spin-coated in a PVA solution on a coverslip and then covered with 4 layers of parafilm (resulting in a $500~{\rm \mu m}$-thick scattering sample) and another coverslip on top (Fig.~\ref{fig:fluo}B). The excitation of the fluorescent beads was achieved by focusing the $532~{\rm nm}$ laser from the rear of the sample with excitation intensity $<10~{\rm kW.cm^{-2}}$. Noteworthy, neither the power of our laser source ($<5~{\rm mW}$) or the single-photon fluorescence-absorption mechanism used, allowed us to efficiently excite single isolated micro-beads through the scattering medium. However, several solutions have been investigated to generate guide-stars in-depth~\cite{horstmeyer2015guidestar}, especially using two-photon excitation~\cite{katz2014noninvasive,ji2017adaptive}. \emph{In fine,} the total fluorescence photon budget at the WFS was $1.4\times 10^8$ photo-electrons, corresponding to a mean photo-electron budget per pixel equal to $33$ (see Fig.~\ref{fig:photon_budget}). The SLM/WFS was conjugated to the achromatic plane of the scattering sample and the fluorescence speckle signal was imaged and measured at the WFS (Fig.~\ref{fig:fluo}C). Finally, the conjugate phase was addressed to the SLM without (Fig.~\ref{fig:fluo}D) and with (Fig.~\ref{fig:fluo}E) including optical vortices, and the playback laser beam was sequentially tuned to both the average fluorescence wavelength ($565~{\rm nm}$) and the excitation wavelength ($532~{\rm nm}$). In both cases, complete phase reconstruction outperforms the regular AO scheme only compensating for ballistic light, further demonstrating the ability to robustly mistune the playback laser wavelength to take into account the fluorophores Stokes' shift at the expense of a slight loss in refocused energy ($14.6\%$ \emph{vs.} $17.9\%$). When conjugating the SLM/WFS to the guide-star plane, no focus could be obtained. 
Noteworthy, the focus quality obtained with fluorescent microbeads (Fig.~\ref{fig:fluo}) are noticeably degraded as compared to all other focii we could obtain in all other conditions, including for spectrally broad or spectrally shifted readout lasers (See Supp.~Fig.~\ref{fig:imagettes}). The focii degradation observed in Fig.~\ref{fig:fluo}D and~\ref{fig:fluo}E is due to the presence of the fluorescent micro-bead on the beam path because its refractive index is not matched with the one of the PVA embedding matrix~\cite{boniface2019noninvasive}. Moreover, it was not possible to shift the playback focus away from the microbead without inducing a important focus-energy degradation because the isoplanatic patch of our sample is typically comparable to bead size itself.
To characterize the DFPC performances, we thus measured the fraction of playback light energy in the main focus as described in Supp.~Fig.~\ref{fig:DOPC_performance_estimation}.
Performances are quantified in Table~\ref{table:DFPC} for three samples thicknesses.

\begin{table*}[!h]
\begin{center}
\begin{tabular}{|c|c|c|c|}
\hline
 & $L=250~{\rm \mu m}$ & $L=375~{\rm \mu m}$ & $L=500~{\rm \mu m}$ \\
\hline
\multicolumn{4}{|c|}{ {\bf DFPC at the average fluorescence wavelength ($\lambda_{rf} = 565{\rm nm}$)}  }\\
\hline
with vortices & $11.3\%\, (\pm5.8)$ & $9.3\%\,(\pm4.1)$ & $7.1\%\,(\pm 6.1)$ \\
without vortices & $6.8\%\, (\pm2.3)$ & $3.9\%\,(\pm0.9)$ & $3.7\%\,(\pm2.1)$\\
\hline
\multicolumn{4}{|c|}{ {\bf DFPC at the excitation wavelength ($\lambda_{rf} = 532{\rm nm}$})}\\
\hline
with vortices & $8.7\%\, (\pm3.3)$ & $9.5\%\,(\pm4.7)$ & $6.5\%\,(\pm 4.8)$ \\
without vortices & $4.7\%\, (\pm1.7)$ & $2.8\%\,(\pm1.5)$ & $2.3\%\,(\pm1.2)$\\
\hline
\end{tabular}
\caption{\label{table:DFPC} DFPC average efficiency ($\pm$ experimental standard deviations) for a playback laser wavelength $\lambda_{rf}$ at the average emission wavelength ($\lambda_{rf}=565\pm2.5 ~{\rm nm}$) and at the excitation wavelength ($\lambda_{rf}=532~{\rm nm}$) from fluorescent light emitted by micro-beads ($\lambda_{GS}=565\pm 15~{\rm nm}$) as a function of paraffin thickness.}
\end{center}
\end{table*}

\section*{Discussion}

In this work, we have demonstrated the ability to perform single-shot DOPC from incoherent guide-stars hidden behind forward scattering samples. This progress has been made possible thanks to two conceptual improvements: the ability to rebuild complex speckle patterns with a high-resolution WFS on the one hand and the large spectral bandwidth revealed by scattering media with large anisotropy factors on the other hand. Our results not only underline the importance of rebuilding optical vortices which have typically been ignored by Shack-Hartmann WFS users but they also prove the key role of associated spiral phase structures for DOPC performances. We also demonstrate the ability to use fluorescent micro-beads as guide-stars which are associated with three issues: the weak fluorescence photon budget challenging the maximum number of modes one may rebuild, the $\simeq 30~{\rm nm}$ spectral bandwidth of the fluorescence signal, and the Stokes' shift between the fluorescence  and the excitation spectra.
The spectral width of DOPC efficiency through $500~{\rm \mu m}$-thick paraffin samples is characterized, and qualitatively matches the spectral correlation width of the medium. The spectral width of the scattering medium is driven by the standard deviation of optical path length trajectories and impose an upper limit for the spectral width of the emission filter. Beyond this limit, the speckle cannot be considered as achromatic and only an averaged wavefront can be defined, wherein optical vortices are scrambled, so reducing the focusing ability of the AO system. In this regard, we also demonstrate the importance of conjugating the WFS and the SLM with the achromatic plane of the scattering slab by demonstrating the degraded performances when conjugation is achieved with the guide-star plane. In the end, the emission filter width is a compromise between the contrast of the fluorescence speckle at the WFS and the photon budget. We estimated that our solution requires an average photon budget per spatial mode of the scattering medium of the order of $ 3\times 10^4$.

In AO systems, the SLM/WFS are often placed in the Fourier plane. Our system did not allow us to easily switch from the sample plane to the Fourier plane and a fair comparison might deserve further experimental characterization.  
As an argument in favor of our configuration, in the Fourier plane, the beam scattered by a forward scattering medium experiences a transverse homothetic dilation for a wavelength shift, so that the maximum number of measurable/controllable modes is limited by $\left(\frac{\lambda}{\delta\lambda}\right)^2$. In contrast, the achromatic plane is the origin of the longitudinal axial dilation responsible for the transverse dilation at infinity~\cite{zhu2020chromato,arjmand2021three}. No such transverse dilation occurs and the speckle correlation is solely limited by the intrinsic spectral correlation width of the scattering medium.
Consequently, for broadband light manipulation, the number of modes we can manipulate is not limited in the achromatic plane, in contrast to the Fourier plane where fluorescence speckles are blurred. This Fourier-conjugated configuration was in particular chosen by the authors of Ref.~\cite{aizik2022fluorescent} where a $10~{\rm nm}$ filter was used to increase the coherence of the fluorescence signal and improve the number of measurable modes. Further quantitative comparison with the technique Ref.~\cite{vellekoop2012dopcFluo} would be deserved, especially regarding the DOPC performances as a function of the nature of phase distortions (including low order aberrations or not) as well as the photon budget per mode.

Noteworthy, provided that our technique enables efficient multiple scattering compensation through samples, in particular those leaving no ballistic light, the guide-star size should not exceed the autocorrelation width of the original speckle pattern. Besides the technical challenge to generate isolated guide-stars, especially in densely labeled samples, the isoplanatic patch size associated with the obtained readout focus appears as critically small for imaging applications. This intrinsic constraint imposed by such samples makes our achievement more relevant to applications such as in-depth photostimulation, thermal phototherapy, in-depth caged-drug photo-delivery, light-induced guided neuronal growth, non-imaging cell activity monitoring or micro-endoscopy through fiber bundles. Brain cells monitoring through the skull also appears as especially interesting thanks to the distance separating the scattering medium (the skull) from the object of interest (the brain), yielding an isoplanatic patch size of $\simeq 15~{\rm \mu m}$~\cite{blochet2021fast,Cui_PNAS_15}.

Beside opening a solution for performing DOPC, this work motivates further fundamental investigations about forward scattering media.
First, we could not attribute the spectral blue-shift observed in Fig.~\ref{fig:detuned_GS} to a mis-calibration of our DOPC system. The difference in the amount of blue-shift between the two sample thicknesses ($L=250~{\rm \mu m}$ and $L=500~{\rm \mu m}$) and between the two imaging conditions (conjugation with the guide-star plane or the achromatic plane) suggests a physical reason for blue-shifted light to yield larger DOPC efficiency than at the guide-star wavelength. Blue-shifted light may benefit more from phase conjugation because of the multiple scattering process scales as $\lambda^2$ in forward scattering media~\cite{arjmand2021three} (in contrast to the $\lambda^4$-scaling in the Rayleigh scattering case~\cite{jacques2013optical}).
Second, by contributing to bridging the gap between AO and wavefront shaping, this work about broadband forward scattering media interrogates about the relevance and the possibility to define a theoretical discrimination criterion between optical aberrations and light scattering. The distortion matrix concept has supported this distinction~\cite{Aubry_SA_20}. Usage has typically used the term aberrations for low-order wavefront distortions only although high-order smooth distortions exhibit no fundamental difference. However, the presence of optical vortices in the wavefront suggests making out from wording aberrations since optical vortices necessarily originate from at least three interfering beams~\cite{Padgett} and thus suppose several optical trajectories.

\section*{Materials and Methods}

\paragraph{Samples preparation.} The $720~{\rm \mu m}$ acute slice of spinal cord of a 6 days mouse was fixed in paraformaldeyde and mounted in fluoromount between two coverslips separated by a $800~{\rm spacer}$. Nail varnish was used to minimize evaporation. The sample was left vertically in place for 24 hours for gravity drifting to stop. 

Parafilm samples were prepared by stacking 2, 3 or 4 layers of parafilms between two coverslips and gently pressed to minimize air spacing and thus surface scattering at interfaces. The thicknesses of the stacks were measured thanks to a calibrated translation stage by observing the back reflexion of the guide-star on the internal coverslip surfaces right aside from the sample. To ensure isotropic scattering, the parafilm layers were oriented at $90^\circ$ for the two-layer samples and at $0^\circ$, $45^\circ$, $90^\circ$ and $-45^\circ$ for the 4-layer samples.

The beads samples were prepared by stacking parafilm diffusers onto a spin-coated layer of beads embeded in poly-vinyl alcool (PVA). The layer was prepared by spin-coating at $10000~{\rm rpm}$ a $10^{-4}$-weight solution of $1~{\rm \mu m}$ fluorescent microbeads (Fluospheres F8820, Thermofisher Scientific) in a $2\%$-weight solution of PVA. For the 2-parafilm-layer sample, two $\#1$-coverslips were used to add a $\simeq 500~{\rm \mu m}$ spacing between the layer of beads and the parafilm.

\begin{figure}
\centering
\fbox{\includegraphics[width=0.97\linewidth]{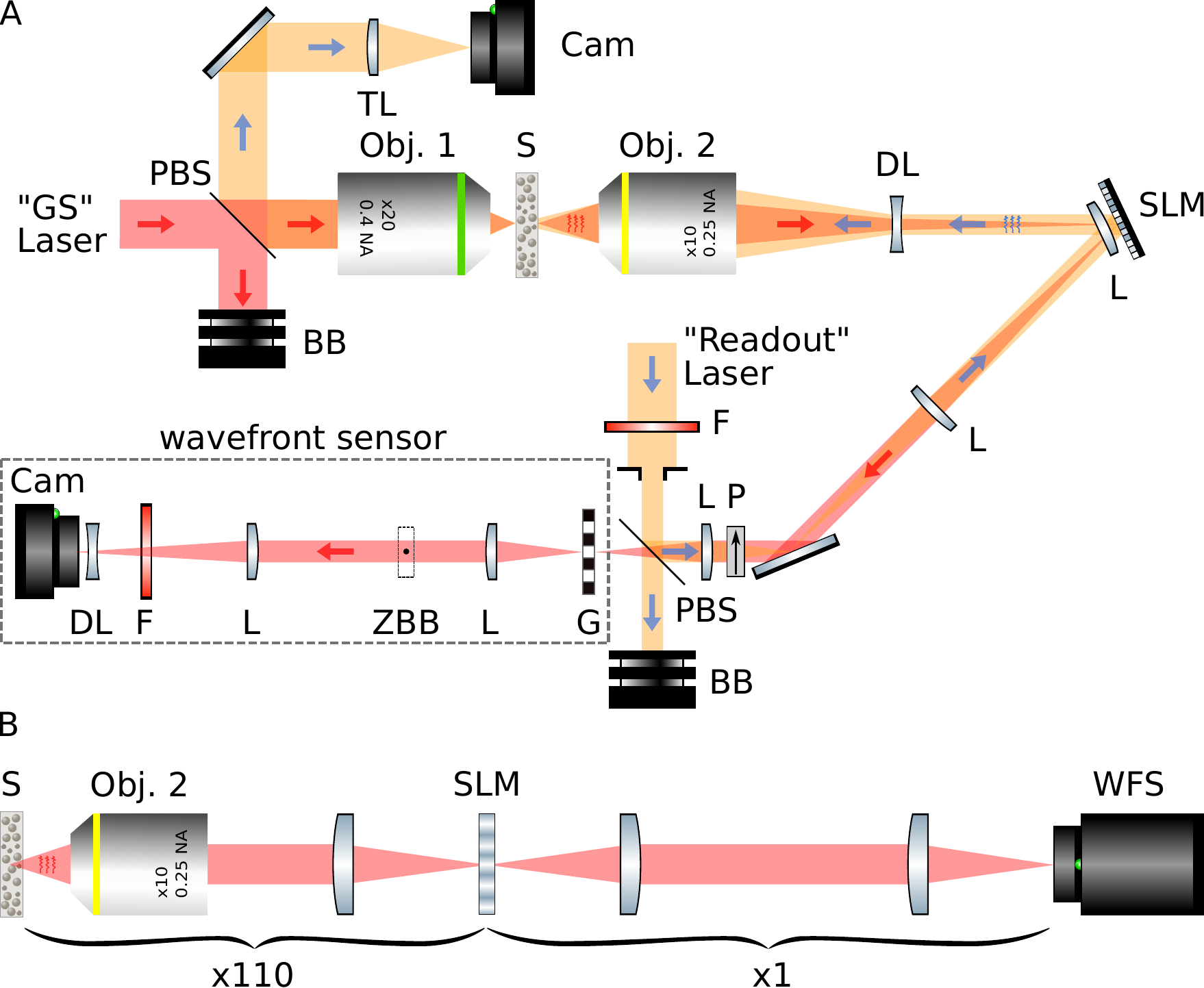}}
\caption{{\bf Full and equivalent experimental scheme of the optical phase conjugating microscope.} A) Full experimental system. PBS: Pellicle beam splitter; BB: Beam block; S: Scattering sample; DL: Diverging lens; L: converging lens; SLM: Spatial light modulator; P: Polarizer; G: grating; ZBB: Zero-order beam block; F: spectral filter used for fluorescence experiment but removed otherwise; Cam: Camera; TL: Tube lens. B) Equivalent representation of the speckle beam control arm. The configuration in A) allows having a large magnification imaging system with reduced size on the optical table.}
\label{fig:setup_DOPC_full}
\end{figure}

\paragraph{Optical setups.}
A detailed description is provided in the supplementaries. In brief, two optical systems were used, a first one for comparing speckle field reconstruction by wavefront sensing and digital holography (see Supp.~Fig.~\ref{fig:setup_WFS_DH}) and a second one to perfom DOPC through scattering samples using a WFS (Supp.~Fig.~\ref{fig:setup_DOPC_full}). To obtain Fig.~\ref{fig:vortex_or_not_vortex}, the first objective lens (Obj.~1) was removed to illuminate the sample with a collimated laser beam. In the DOPC experiments, the SLM and WFS were conjugated to the so-called ``achromatic plane'', located in a virtual plane lying at a depth equal to $2/3$ of the sample thickness (or equivalently, the fluorescent guide-start depth). The location of the output plane of the scattering sample was first identified thanks to white light illumination and the objective was then moved forward of the right amount towards the scattering sample.

In the DOPC experiment, the guide-star was generated by using a focused laser beam originating from fiber-coupled diode laser ($532~{\rm nm}$, $635~{\rm nm}$) or a super-continuum laser (ElectroVis-470, LEUKOS, France) combined to a computer-tunable filter box (Bebop, LEUKOS, France). The laser was focused into the sample with a $\times 20$, $0.4~{\rm NA}$ objective lens and the scattered light was collected using a $\times 10$, $0.25~{\rm NA}$ objective lens. 

In the fluorescent guide-stars experiment in Fig.~\ref{fig:fluo}, the microbeads were excited from the rear of the scattering medium using a focused $532~\rm nm$ laser. The excitation intensity was $<10~{\rm kW.cm^{-2}}$ and the exposure time at the WFS was $1~{\rm s}$.

The WFS design is similar to the one described in Ref.~\cite{Bon_NM_18}. It is made of a two-dimensional $20~{\rm \mu m}$-step checkerboard $0-\pi$-phase grating conjugated to a CMOS camera (BSI Prime Express, Photometrics) thanks to a relay telescope. An additional $-75~{\rm mm}$ diverging lens was fixed at the C-mount thread of the camera to minimize field curvature introduced by the relay telescope. The final global magnification of the imaging-relay system was $\simeq 3$. 

In the first optical system comparing digital holography to wavefront sensing (Supp.~Fig.~\ref{fig:setup_WFS_DH}), a $635~{\rm nm}$ diode laser was spatially filtered by a monomode fiber and collimated to illuminate a $1^\circ$ holographic diffuser (Edmund Optic). This diffuser was tuned out-of-focus the WFS and an iris was placed in an intermediate Fourier plane to tune the speckle grain size. For DH experiments, the grating of the WFS was removed and a reference arm generated from a beam splitter placed before the scattering sample. The reference beam and the ``signal'' speckle beam were combined together with an angle larger than the opening angle of the speckle, thanks to a pellicle beam-splitter located in the Fourier plane of the relay telescope of the WFS.

Finally, a polarizer was added in the optical path of the WFS because scattering media, even forward scattering, slightly depolarize light beams~\cite{Alfano_PRL_05}. In our case, we measured that, for a thickness $L=500~{\rm \mu m}$ of paraffin, the polarization rate was $0.83$. 
This slight depolarization thus accounts for $\simeq 17\%$ of degradation in DOPC performances since our SLM can only modulate a single polarization component. 

\section*{Acknowledgments}
The authors thank Boris Lamotte D'Incamps and Brigitte Delhomme for the preparation of biological specimens. MG acknowledges Sylvain Gigan for encouraging and stimulating discussions.

\section*{Funding}
This work was partially funded by the technology transfer company SATT ERGANEO (Project 600) and the French Research National Agency (projects SpeckleSTED ANR-18-CE42-0008-01 and MaxPhase ANR-20-CE42-0006). MG acknowledges support for Institut Universitaire de France.



\clearpage

\beginsupplement

\twocolumn[{%
 \centering
 \bigskip
 \huge Supplementary Materials\\[0.5em]
 \Large Single-shot Digital Optical Fluorescence Phase Conjugation \\ Through Forward Multiply Scattering Samples\\[3em]
}]

\section{Quantitative measurement of complex field with a wavefront sensor}
\subsection{Experimental setup}
\label{subsection:experimental_setup_WFS_DH}
\begin{figure}[h!]
\centering
\fbox{\includegraphics[width=0.97\linewidth]{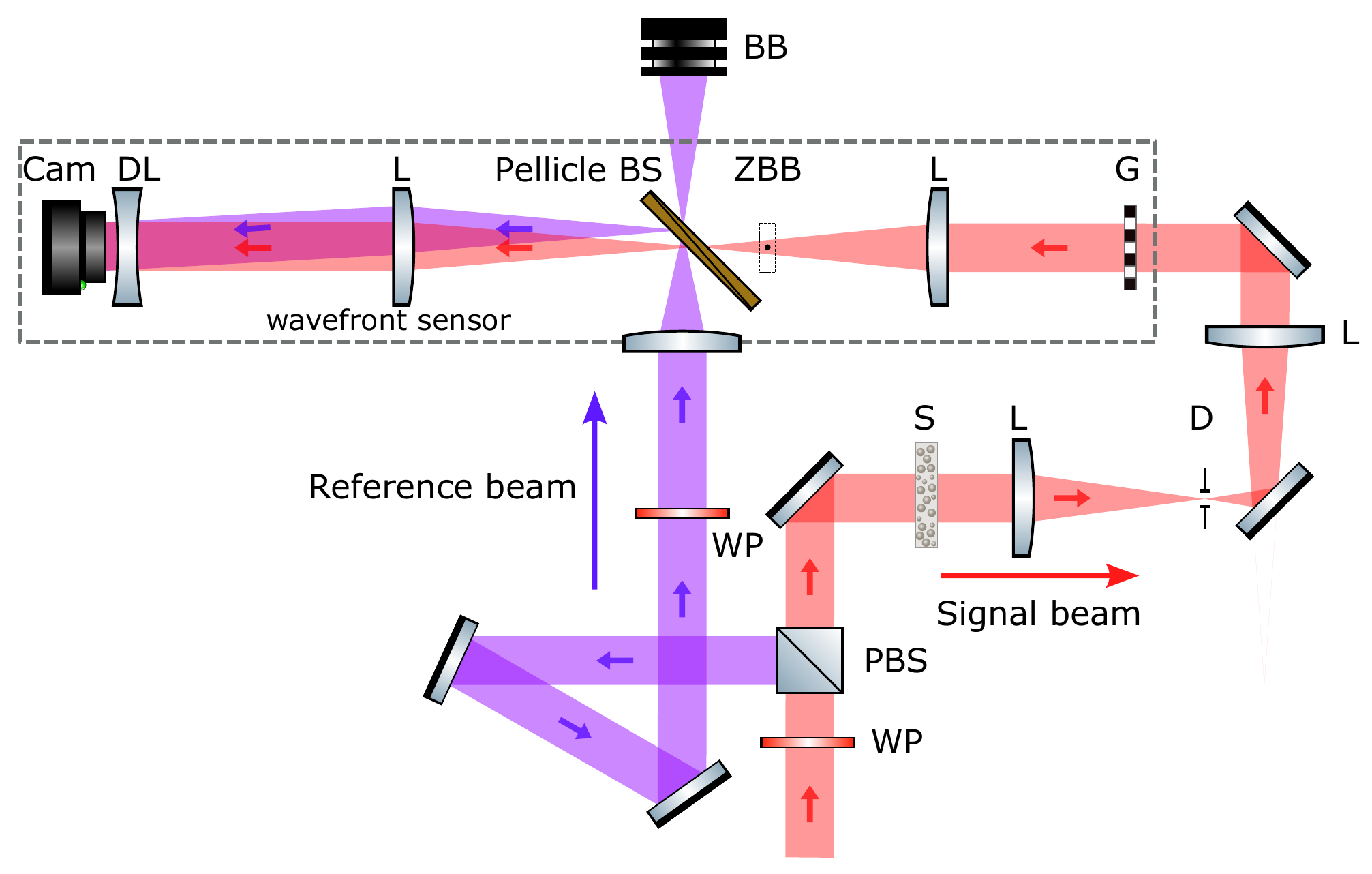}}
\caption{{\bf Full experimental setup of the demonstration of the quantitative measurement of complex field by a lateral shearing interferometer WFS.} WP: Half wave plater; PBS: Polarized beam splitter; S: Scattering sample; L: Lens; D: Diaphragm; G: Grating; ZBB: zero-order beam blocker; BB: Beam blocker; Pellicle BS: Pellicle beam splitter; DL: Divergence lens; Cam: Camera}
\label{fig:setup_WFS_DH}
\end{figure}

\begin{figure}
\centering
\fbox{\includegraphics[width=0.97\linewidth]{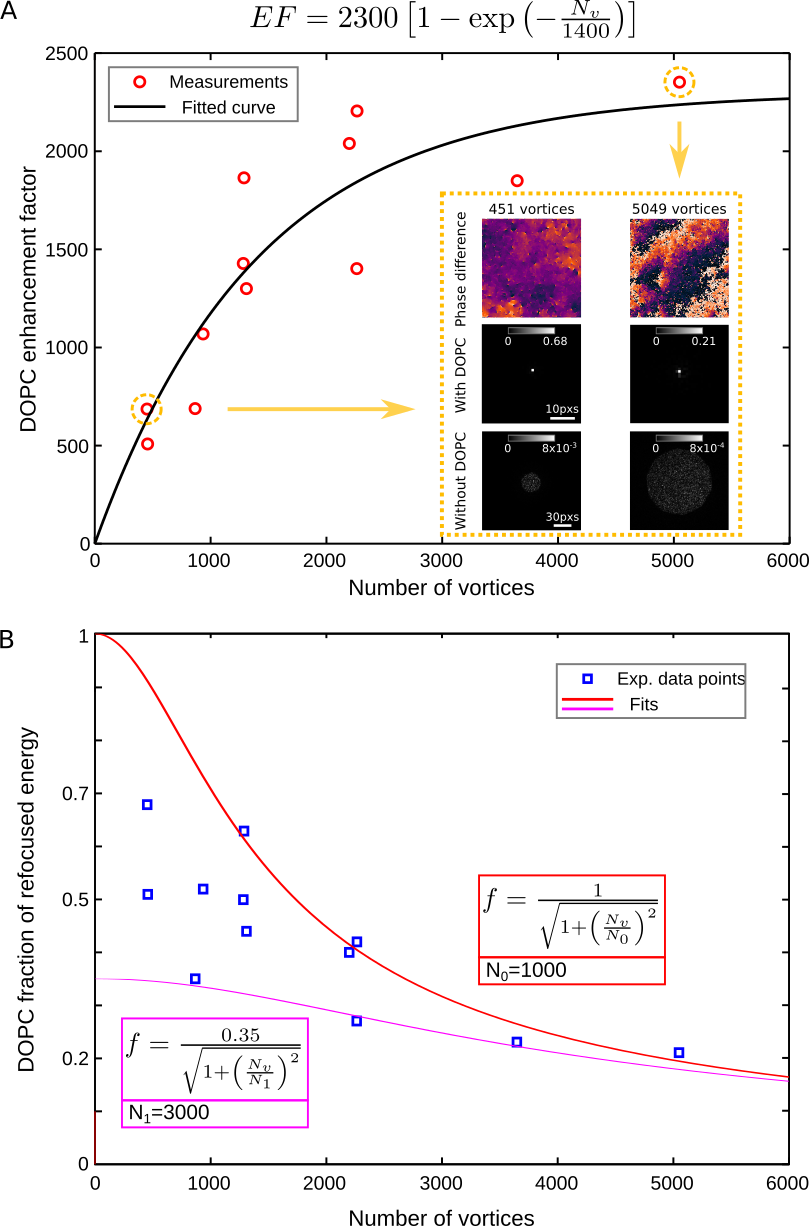}}
\caption{{\bf Experimental estimation of the maximum number of modes that our WFS can rebuild to achieve DOPC.} The number of vortices (\emph{i.e.} the number of modes) is varied by tuning an iris diaphragm in a Fourier plane (D in Fig.~\ref{fig:setup_WFS_DH}). The phase is measured with the WFS and compared to the phase measured by DH, the latter phase pattern being considered as the ground truth. Then, digital optical phase conjugation (DOPC) is numerically simulated based on the phase difference, both to compute the enhancement factor (A) and the fraction of refocused energy (B). Experimental data points are fitted using empirical curves to underline and quantify the global trend. As expected, the slope of the DOPC for small number of modes is close to one~\cite{vellekoop2007focusing,jang2017optical}.}
\label{fig:WFS_DH_number_of_modes}
\end{figure}

In this section, we experimentally demonstrate that our wavefront sensor (WFS) can quantitatively measure a scattered speckle field from a single-shot acquisition by comparing the retreived phase and amplitudes to the one obtained by off-axis digital holography (DH), which we consider as a ground truth. Fig.~\ref{fig:setup_WFS_DH} shows the full experimental setup, in which the collimated beam with a wavelength of $635~{\rm nm}$ passes through a polarized beam splitter, separated into the reference beam path and the signal beam path. In the signal beam path, the beam interacts with a $5^\circ$ holographic diffuser (Edmund Optic), and the complex field is imaged on the WFS by a telescope with a lens set of f=$100~{\rm mm}$ and f=$200~{\rm mm}$, respectively. In order to tune the speckle density, an adjustable diaphragm is placed in the intermediate Fourier plane of the telescope. The WFS is composed by a two-dimensional $20~{\rm \mu m}$-step checkerboard phase grating, placed $3~{\rm mm}$ away from the camera sensor, a telescope with a lens set of f=$100~{\rm mm}$ and f=$200~{\rm mm}$, and a divergence lens with a focal length f=$-50~{\rm mm}$, which is used to correct the field curvature. A beam block (ZBB) is placed closed to the intermediate Fourier plane of the telescope in the WFS, to only select the first diffracted order of the grating pattern (blocking the zero order and higher orders). In the reference beam path, a set of mirrors are used to adjust the length of the optical path and the tilt angle between two beams to optimize the contrast of interference fringes in DH measurement. A lens with a focal length f=$50~{\rm mm}$ and a pellicle beam splitter is used to image the reference beam on the camera. Two half waveplates are placed respectively in the two paths to adjust the relative beam intensity. A beam blocker BB is used to collect the useless beam in the experiments. In the data acquisition step, for both methods we take two measurements with and without scattering sample and compare the relative complex field between the two measurements for each method. In order to be sure that the two methods measure the same scattering signal, we first measure the field without scattering sample in DH (i.e. open both the two paths and remove the grating G and beam blocker ZBB). Then we insert the scattering sample and measure the scattering field in DH. After finishing the measurement in DH, the reference beam path is blocked and we place the grating G and beam blocker ZBB to do the measurement with scattering sample in WFS. Finally, we remove the scattering sample to do another measurement in WFS. 50 frames are recorded and averaged for each measurement for better signal to noise ratio.   

\subsection{Phase gradient measurement from a lateral shearing interferometer WFS}
\begin{figure}
\centering
\fbox{\includegraphics[height=1.33\linewidth]{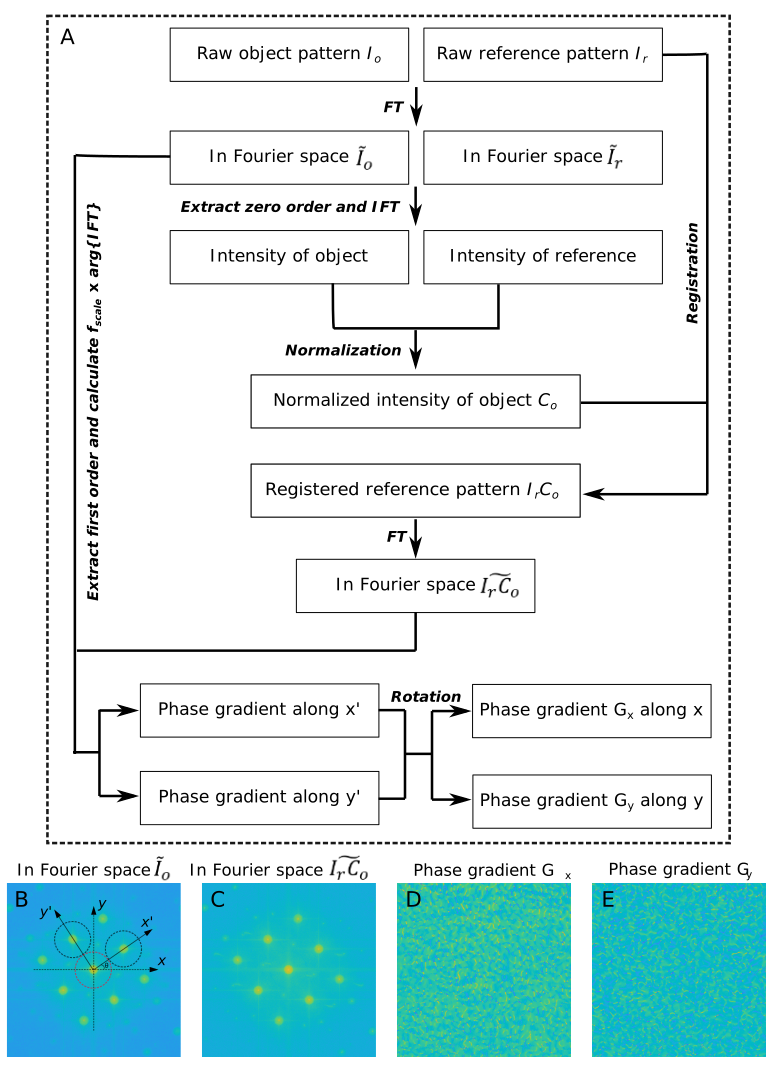}}
\caption{{\bf Phase gradient reconstruction procedure from a quadri-wave lateral shearing interferometer WFS.} 
}
\label{fig:WFS_PhaGrad}
\end{figure}

Off-axis DH can directly measure a complex field (phase and amplitude) by extracting the first order of the fringe pattern in the Fourier space. In contrast, the WFS does not measure directly the phase but the phase gradient. 
Here we describe in detail the procedures to extract the phase gradient information from the raw camera image of a quadriwave lateral shearing interferometer (QWLSI)~\cite{primot1995achromatic,bon2009quadriwave}. A detailed description of the full gradient recovery process  from QWLSI measurements is given in Ref.~\cite{Baffou_JPD_21}. Fig.~\ref{fig:WFS_PhaGrad} briefly summarizes and illustrates the main steps. 

\subsection{Measurement of complex scattering field WFS vs DH}
\begin{figure}
\centering
\fbox{\includegraphics[width=0.97\linewidth]{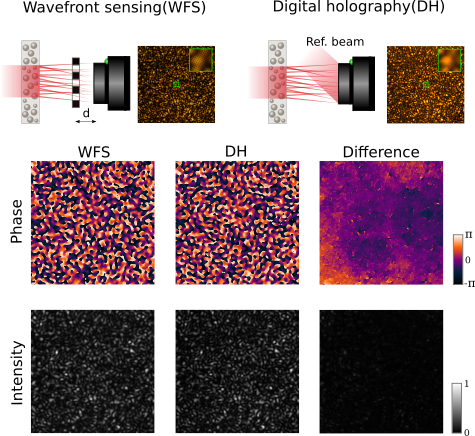}}
\caption{{\bf Comparison of speckle field reconstruction using a lateral shearing interferometer WFS and off-axis digital holography (DH).}}
\label{fig:comparison_WFS_DH}
\end{figure}

In Fig.~\ref{fig:comparison_WFS_DH}, we compare the complex field reconstruction obtained from both the WFS and off-axis DH, thanks to the optical system shown in Fig.~\ref{fig:setup_WFS_DH}. In these images, the number of vortices in the beam at the WFS (\emph{i.e.} the number of spatial modes) was $1581$. The intensity image in the DH part is obtained by blocking the reference beam for direct intensity recording. Both the phase difference and intensity difference are shown, demonstrating the good agreement between WFS and DH measurements. A remaining low-spatial frequency content appears in the phase difference pattern, probably due to cumulative integration errors of phase gradients estimates over the large number of modes.

\section{Experimental setup and calibration for digital optical phase conjugation with a wavefront sensor}
\subsection{Experimental setup}
\label{subsection:experimental_setup_DOPC}
In this section, we mainly show the detailed experimental setup for performing the digital optical phase conjugation (DOPC) with a WFS. A collimated guide-star (GS) laser beam passes through a pellicle beam splitter (PBS) and an objective Obj.1 (x20, 0.4NA) focuses the beam that generates a guide-star (or excites a fluorescent micro-bead guide-star) at the rear of the scattering sample S. The scattered field is imaged onto the SLM by an afocal telescope, which is composed by an objective Obj.2 (x10 0.25NA), a diverging lens DL with a focal length f=$-50~{\rm mm}$ and a wavefront correcting lens with a focal length f=$1000~{\rm mm}$ placed in front of the SLM. As compared to a regular Galileo telescope made of two converging lenses in a 4-f configuration, our system has the advantage to allow large magnification ($\times 110$) between the sample plane and the SLM plane over a shortened physical distance. Then, a telescope with a lens set of f=$300~{\rm mm}$ and f=$150~{\rm mm}$ is used to conjugate the SLM plane to the WFS, which has the same components as the previous experimental setup described in Sec.\ref{subsection:experimental_setup_WFS_DH}. A removable mirror (not shown) can be placed between the pellicle beam splitter and the WFS grating G to enable the recording of the reference pattern. After measuring the scattered field, the conjugated phase is loaded on the SLM. A collimated "playback" laser beam is reflected by a polarized beam splitter and propagates back to the SLM. The linear polarizer P, used to polarize the beam originating from the scattering sample is also used to clean the polarization of the "playback" laser beam and match the modulating polarization axis of the liquid-crystal SLM. The wavefront modulated by the SLM then passes through the scattering sample. The focus of the "playback" laser beam at the guide-star position is monitored on a camera through Obj.1 thanks to a tube lens TL (f=$150~{\rm mm}$).

\subsection{Optimization of the distance between the grating and the camera}
\begin{figure}
\centering
\fbox{\includegraphics[height=1.33\linewidth]{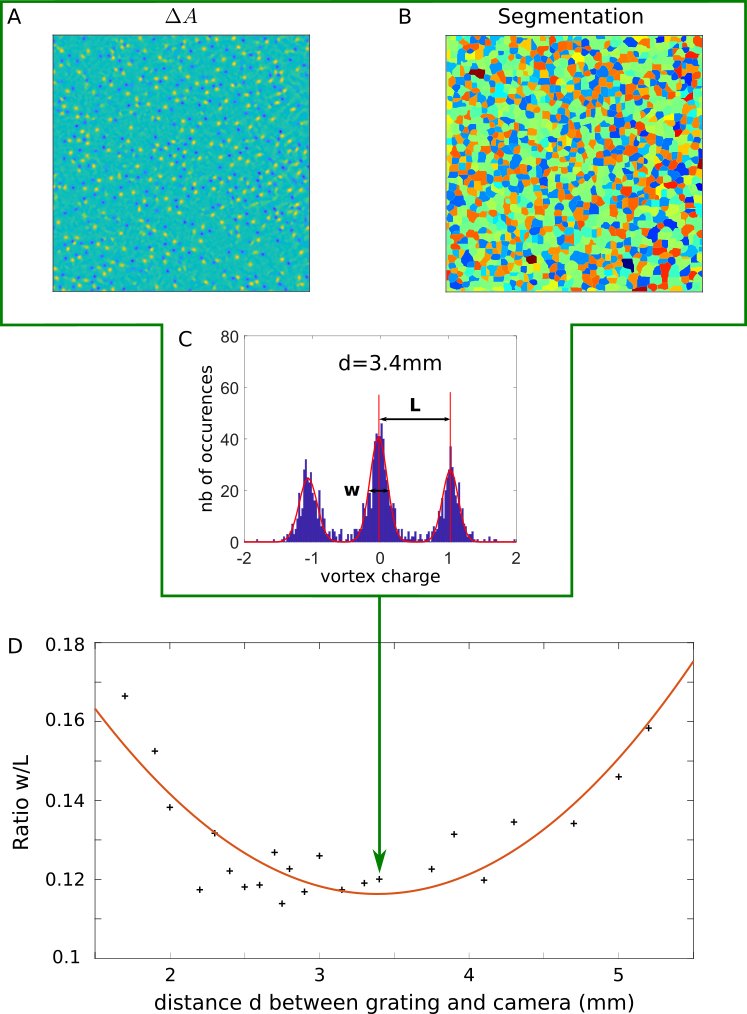}}
\caption{{\bf Optimization of the distance grating-camera in the WFS for speckle patterns measurement.} (A) Laplacian of the vector potential shows the location of the optical vortices in a speckle field. (B) Image segmentation identifies the optical vortices map for topological charge measurement. (C) Histogram of the measured charges of optical vortices. The optimized grating-camera distance can be determined by the contrast of the histogram, defined as the ratio $w/L$ where $w$ is the peak width and $L$ the distance between peaks. (D) Plot of the measured contrast as a function of grating-camera distance.
}
\label{fig:optimization_distance_d}
\end{figure}

For a lateral shearing interferometer WFS, the grating-camera distance $\emph{d}$ drives a trade-off between the phase sensitivity, phase gradient dynamic and spatial resolution. In our case, spatial resolution is of critical importance to measure complex wavefields containing as many spatial modes as possible. The number of modes roughly equals the number of optical vortices of a fully developed speckle pattern. Fig.~\ref{fig:optimization_distance_d}A shows the optical vortex distribution in a measured complex scattering field. The yellow and blue peaks denote the random locations of vortices of charges $+1$ and $-1$, respectively. $\Delta A$ stands for the Laplacian of the vector potential $\emph{A}$, which account for the solenoidal contribution in the Helmholtz decomposition of the phase gradient vector field measured at the WFS, as detailed in Ref.~\cite{wu2021reference}. $\Delta A$ yields peaks (in theory Dirac functions) at vortices locations. Segmentation splits the $\Delta A$ image into multiple small regions according to the locations of the peaks. A following integration within each small region gives the specific vortex charge in each location, according to Stokes’ theorem~\cite{wu2021reference}. The histogram of the distribution of the estimated charges of the optical vortices in the complex scattering field is shown in Fig.~\ref{fig:optimization_distance_d}C, in which three peaks occurs at charge +1/-1 and 0. Here we use the histogram contrast, which is defined by the ratio between the width of the central peak $\emph{w}$ and the distance $\emph{L}$ between the central peak and the side peak, to evaluate the vortex-discrimination ability of the WFS. The ratio $\frac{w}{L}$ is then plotted as a function of the distance $\emph{d}$. The optimal distance $\emph{d}$ is finally obtained by computing the minimum coordinate of a parabolic fitting curve. 

\subsection{Alignement and calibation of SLM and WFS}
\begin{figure}
\centering
\fbox{\includegraphics[width=0.97\linewidth]{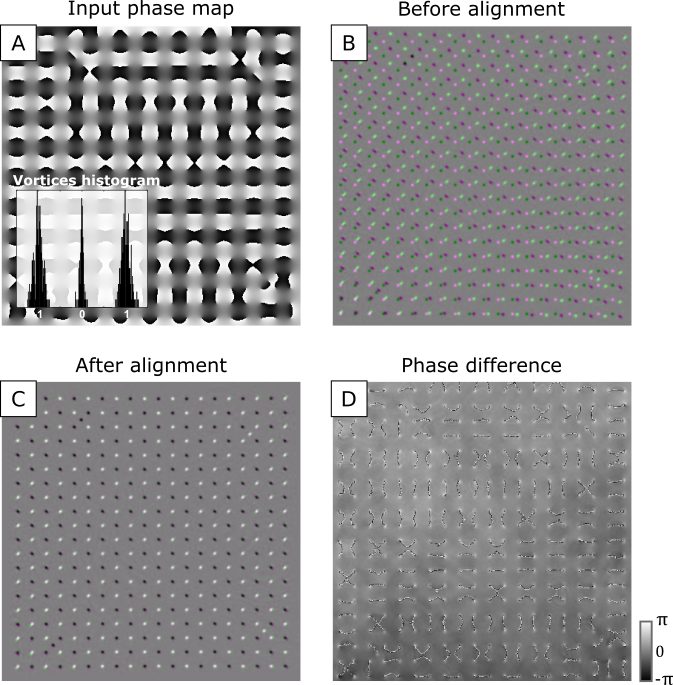}}
\caption{{\bf Alignement and calibation of the SLM and WFS.} (A) Input phase map generated by a specific vortices array, used to align the mismatch between the SLM and the WFS plane. Inset is the histogram of the measured optical vortices. (B) Merged vortices maps of the input and the direct measurement from the WFS before alignment. (C) Merged vortices maps of the input and the measurement after alignment. (D) Phase difference between the input phase map and the measured phase map with alignment.}
\label{fig:alignment_SLM_WFS}
\end{figure}

In order to compensate the scattering field in the DOPC experiments, a proper alignment and calibration of SLM and WFS is highly required. Optical elements conjugating the SLM to the WFS may induce some geometric transformation such as translation, rotation, and dilation between the SLM and WFS plane. In our experiments, we calibrate alignment issues thanks to a custom-designed vortex array. Specifically, we design a 20x20 vortex array with alternating $+1$ and $-1$ charges. Four additional symmetry-breaking vortices were added to the array to eliminate symmetry ambiguity problems. The grid thus contain $404$ optical vortices in total. A phase map corresponding to this optical vortices array is shown in Fig.~\ref{fig:alignment_SLM_WFS}A. This pattern is loaded on the sub-region of the SLM lying in the field of view of the WFS. A collimated beam then illuminates the SLM and the reflected beam is then measured by the WFS. The $\Delta A$ map is then computed and compared to the generated vortex pattern. The histogram of the measured optical vortices is shown as an inset in Fig.~\ref{fig:alignment_SLM_WFS}A. The calibration of the SLM is achieved by centering the peaks of the vortex charge histogram on $\pm 1$ values. A typical comparison between a still non-aligned measured vortex location and the loaded true vortices array is shown in Fig.~\ref{fig:alignment_SLM_WFS}B. To correct this mismatch, we evaluate the geometric transformation of the two vortices arrays by using the built-in function in Matlab ("imregtform", Matlab Image Processing Toolbox). With the evaluated transformation coefficients, we introduce a numerical correcting transform to the pattern measured by the WFS to align the two patterns as shown in Fig.~\ref{fig:alignment_SLM_WFS}C. We then compute the phase difference between the loaded phase map (in Fig.~\ref{fig:alignment_SLM_WFS}A) and the phase map measured by the WFS in Fig.~\ref{fig:alignment_SLM_WFS}D, where all phase vortices perfectly cancel out. This flat phase pattern ensures the possibility to achieve phase conjugation.

\section{Combination of DOPC and computer generated holography}
\begin{figure}
\centering
\fbox{\includegraphics[width=0.97\linewidth]{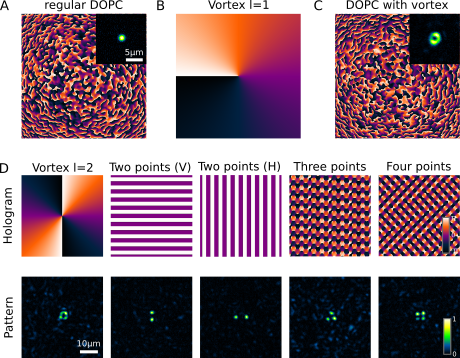}}
\caption{{\bf Computer generated holography combined with DOPC for patterned excitation.} (A) The phase map used to perform DOPC through a 1$^\circ$ holographic diffuser. Inset is the focus after DOPC. (B) Hologram pattern for generating +1 charge optical vortex. (C) The resulted phase map for delivering a donut (+1 charge information) through the scattering sample. Inset is the result with DOPC. (D) Hologram patterns and corresponding focused patterns through the diffuser.}
\label{fig:hologram_delivery}
\end{figure}

In this section, we demonstrate two more DOPC related experiments with the measurement of the complex scattering field by WFS. We show that as long as the scattering is properly compensated, one can deliver a specifically designed hologram through the scattering sample. The experimental setup presented in Sec.\ref{subsection:experimental_setup_DOPC} is used. The scattering sample here is a 1$^\circ$ holographic diffuser (Edmund Optics), ensuring a sufficient memory effect for the hologram delivery. The collimated beam with a wavelength of $635~{\rm nm}$ passes through the object Obj. 1 and forms a virtual guide-star, locating 3mm away from the diffuser surface. The SLM and WFS are conjugated to the guide-star plane. After measuring the complex scattering field, DOPC is performed through the scattering sample, as is shown in Fig.~\ref{fig:hologram_delivery}A the phase map and the refocus. In the next step, by adding another $-1$-charged optical vortex phase map (Fig.~\ref{fig:hologram_delivery}B) into the measured DOPC phase pattern, instead of only having a focus, we obtain a donut-shaped optical vortex beam on the camera, as is shown in (Fig.~\ref{fig:hologram_delivery}C). Similarly, other hologram patterns can also be delivered through the scattering sample with the measured complex scattering field. Several computer generated holograms could then be generated using a Gerchberg–Saxton algorithm~\cite{Yang_JNE_11}. The delivery results are respectively shown in Fig.~\ref{fig:hologram_delivery}D. Such hologram pattern delivery can be interesting for e.g. the application of photostimulation. 

\section{DOPC through a multimode fiber}
\begin{figure}
\centering
\fbox{\includegraphics[width=0.97\linewidth]{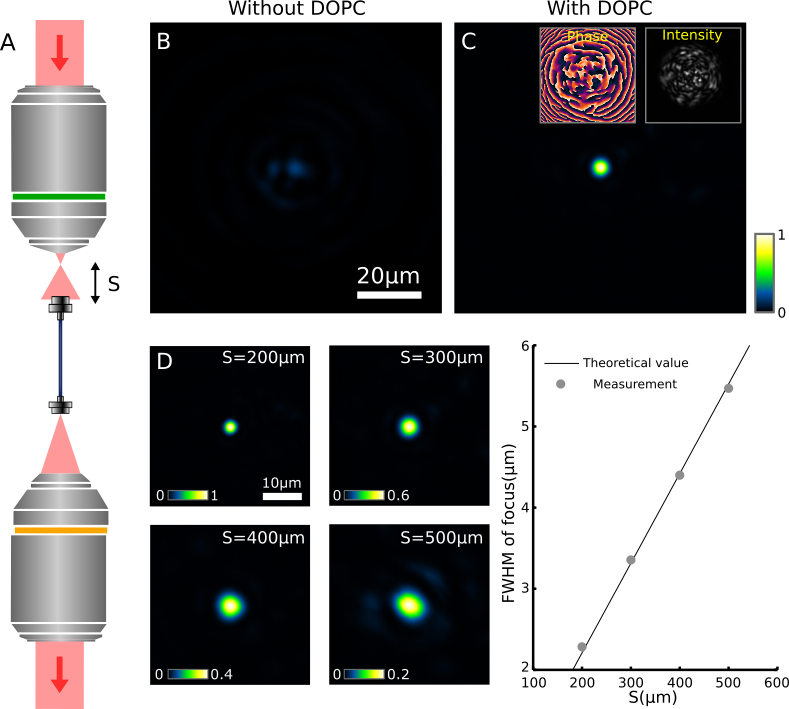}}
\caption{{\bf DOPC through a multimode fiber.} (A) Schematic of the experiments. (B) The recorded image on the monitoring camera without DOPC. (C) DOPC results at S = $300~{\rm \mu m}$ with the complex field measured by WFS. Insets are respectively the phase and intensity. (D) DOPC results for various distances of S and the plot of the size of the focus as a function of S.}
\label{fig:DOPC_fiber}
\end{figure}

This section aims at demonstrating the potential of our DOPC approach to refocus a laser beam through a multimode fiber. Specifically, we use a $1~{\rm cm}$ long multimode fiber (FG050UGA, thorlabs), and the virtual guide-star is generated at S = $300~{\rm \mu m}$ away from the input fiber facet, as is shown in Fig.~\ref{fig:DOPC_fiber}A. The output facet is conjugated to the SLM and WFS and the complex field is measured by WFS. We show in Fig.~\ref{fig:DOPC_fiber}B the playback camera image with flat phase displayed at the SLM, showing a speckle intensity. The image after DOPC is compared in Fig.~\ref{fig:DOPC_fiber}C, which gives a more intense focus on the camera (same colorbar). The complex field measured by the WFS is shown in Fig.~\ref{fig:DOPC_fiber}C as insets. By translating the guide-star to different axial positions from S = $200~{\rm \mu m}$ to S = $500~{\rm \mu m}$, we implement DOPC on all the cases, as we show in Fig.~\ref{fig:DOPC_fiber}D. We can notice that the peak intensity of the focus decreases with increase of S, while the size of the focus (defined as the full-width- half-maximum) gets larger, and it follows the relation of $\sim \frac{\lambda{S}}{D}$, which is shown as a fitting plot in Fig.~\ref{fig:DOPC_fiber}D.

\end{document}